\RequirePackage[mathlines,running]{lineno}
\documentclass[twocolumn,trackchanges]{aastex63}

\usepackage[normalem]{ulem}

\usepackage{lineno}
\usepackage{amssymb}
\usepackage{amsmath,array,multirow}
\usepackage{comment}
\usepackage{enumerate}
\usepackage{bm}

\usepackage{multirow}

\usepackage[nolist,nohyperlinks]{acronym}

\begin{document}

\newcommand{\secref}[1]{\S\ref{#1}}

\newcommand{\bse}{\ensuremath{\texttt{BSE}}}
\newcommand{\emcee}{\ensuremath{\texttt{emcee}}}
\newcommand{\psycris}{\ensuremath{\texttt{psy-cris}}}
\newcommand{\posydon}{\ensuremath{\texttt{POSYDON}}}
\newcommand{\mesa}{\ensuremath{\texttt{MESA}}}

\newcommand{\CIERA}{Center for Interdisciplinary Exploration and Research in Astrophysics (CIERA), 1800 Sherman, Evanston, IL 60201, USA}
\newcommand{\NUPhysAstro}{Department of Physics \& Astronomy, Northwestern University, 2145 Sheridan Road, Evanston, IL 60208, USA}

\newcommand{\gyr}{{\rm{Gyr}}}
\newcommand{\myr}{{\rm{Myr}}}
\newcommand{\msun}{{M_\odot}}
\newcommand{\rsun}{{R_\odot}}
\newcommand{\lsun}{{L_\odot}}
\newcommand{\au}{{\rm{AU}}}
\newcommand{\pc}{{\rm{pc}}}
\newcommand{\henon}{{H\'enon}}
\newcommand{\metal}{{\rm{Z}}}
\newcommand{\metalsun}{{\rm{Z}_\odot}}
\newcommand{\kpc}{{\rm{kpc}}}
\newcommand{\rcobs}{{r_{c,\rm{obs}}}}
\newcommand{\rhl}{{r_{\rm{hl}}}}
\newcommand{\Sigmacobs}{{\Sigma_{c,\rm{obs}}}}
\newcommand{\popone}{{\tt{Pop1}}}
\newcommand{\poptwo}{{\tt{Pop2}}}
\newcommand{\Lto}{{L_{\rm{cut}}}}
\newcommand{\Deltarfifty}{{\Delta_{r50}}}
\newcommand{\Deltaa}{{\Delta_A}}
\newcommand{\nbh}{{N_{\rm{BH}}}}
\newcommand{\ncluster}{{N_{\rm{cluster}}}}
\newcommand{\rlim}{{r_{\rm{lim}}}}
\newcommand{\nrec}{{n_{\rm{rec}}}}
\newcommand{\ninject}{{n_{\rm{inj}}}}
\newcommand{\pmsigma}{\pm 1\sigma}
\newcommand{\gaia}{{\it Gaia}}
\newcommand{\cosmic}{{\texttt{COSMIC}}}
\newcommand{\kps}{{{\rm km\ s}^{-1}}} 
\newcommand{\days}{\rm{day}}
\newcommand{\yr}{\rm{yr}}
\newcommand{\ms}{\rm{ms}}
\newcommand{\logg}{$\rm{log\,g}$}
\newcommand{\TESS}{{\it{TESS}}}
\newcommand{\swift}{{\it Swift}}
\newcommand{\mrem}{{M_{\rm{rem}}}}
\newcommand{\mbh}{{M_{\rm{BH}}}}
\newcommand{\porb}{{P_{\rm{orb}}}}
\newcommand{\teff}{{T_{\rm{eff}}}}
\newcommand{\mgs}{{M_{\rm{GS}}}}
\newcommand{\Lx}{{L_{\rm{X}}}}
\newcommand{\Fx}{{F_{\rm{X}}}}
\newcommand{\mas}{\rm{mas}}
\newcommand{\Porb}{\ensuremath{P_{\rm orb}}}
\newcommand{\Msun}{\ensuremath{{M_\odot}}}
\newcommand{\Mtot}{\ensuremath{{M_{\rm tot}}}}
\newcommand{\RV}{\ifmmode {{\rm RV}}\else RV \fi}
\newcommand{\bigG}{\ensuremath{\mathcal{G}}}

\newcommand{\rot}{ \ensuremath{{\omega/\omega_\mathrm{crit}}} }
\newcommand{\adot}{ \ensuremath{{\dot{a}_\mathrm{eMT}}} }
\newcommand{\edot}{ \ensuremath{{\dot{e}_\mathrm{eMT}}} }

\begin{acronym}[MPC]
\acro{BBH}{binary black hole}
\acro{BH}{black hole}
\acro{NS}{neutron star}
\acroplural{NS}[NSs]{neutron stars}
\acro{WD}{white dwarf}
\acroplural{WD}[WDs]{white dwarfs}

\acro{CO}{compact object}
\acroplural{CO}[COs]{compact objects}
\acro{BPS}{Binary Population Synthesis}

\acro{ZAMS}{Zero Age Main Sequence}
\acro{TAMS}{Terminal Age Main Sequence}
\acro{MS}{main sequence}

\acro{SFH}{star formation history}

\acro{CE}{common envelope}
\acro{SN}{supernova}
\acroplural{SN}[SNe]{supernovae}
\acro{RLO}{Roche-lobe overflow}
\acro{MT}{mass transfer}

\acro{eMT}{eccentric mass transfer}
\acro{s-eMT}{simplified eccentric mass transfer}

\acro{f-eMT}{full eccentric mass transfer}

\acro{HMXB}{high-mass X-ray binary}
\acroplural{HMXB}[HMXBs]{high-mass X-ray binaries}
\acro{BeXB}[BeXB]{Be X-ray binary}
\acroplural{BeXB}[BeXBs]{Be X-ray binaries}

\acro{AL}{active learning}
\acro{ML}{machine learning}
\acro{RBF}{radial basis function}
\acro{PTMCMC}{parallel tempered Markov-chain Monte Carlo}

\acro{mesa}[\ensuremath{\texttt{MESA}}]{Modules for Experiments in Stellar Astrophysics}
\acro{posydon}[\ensuremath{\texttt{POSYDON}}]{POpulation SYnthesis  with Detailed binary-evolution simulatiONs}
\acro{bse}[\ensuremath{\texttt{BSE}}]{Binary Stellar Evolution}
\acro{sse}[SSE]{single star evolution}

\acro{3D}{three-dimensional}
\acro{1D}{one-dimensional}
\end{acronym}

\title{Mass Transfer in Eccentric Orbits with Self-consistent Stellar Evolution}

\author[0000-0003-4474-6528]{Kyle~Akira~Rocha}
\affiliation{\CIERA{}}
\affiliation{\NUPhysAstro{}}
\affiliation{NSF-Simons AI Institute for the Sky (SkAI), 172 E. Chestnut St., Chicago, IL 60611, USA}
\email{kylerocha2024@u.northwestern.edu}

\author{Rachel~Hur}
\affiliation{\CIERA{}}
\affiliation{The Department of Physics, 5720 South Ellis Avenue, Chicago, IL 60637}

\author[0000-0001-9236-5469]{Vicky~Kalogera}
\affiliation{\CIERA{}}
\affiliation{\NUPhysAstro{}}
\affiliation{NSF-Simons AI Institute for the Sky (SkAI), 172 E. Chestnut St., Chicago, IL 60611, USA}

\author[0000-0001-6692-6410]{Seth~Gossage}
\affiliation{\CIERA{}}
\affiliation{NSF-Simons AI Institute for the Sky (SkAI), 172 E. Chestnut St., Chicago, IL 60611, USA}

\author[0000-0001-9037-6180]{Meng~Sun}
\affiliation{\CIERA{}}

\author[0000-0002-2077-4914]{Zoheyr Doctor}
\affiliation{\CIERA{}}

\author[0000-0001-5261-3923]{Jeff~J.~Andrews}
\affiliation{Department of Physics, University of Florida, 2001 Museum Rd, Gainesville, FL 32611, USA}
\affiliation{Institute for Fundamental Theory, 2001 Museum Rd, Gainesville, FL 32611, USA}


\author[0000-0002-3439-0321]{Simone~S.~Bavera}
\affiliation{Département d’Astronomie, Université de Genève, Chemin Pegasi 51, CH-1290 Versoix, Switzerland}
\affiliation{Gravitational Wave Science Center (GWSC), Université de Genève, CH1211 Geneva, Switzerland}

\author[0000-0002-6842-3021]{Max~Briel}
\affiliation{Département d’Astronomie, Université de Genève, Chemin Pegasi 51, CH-1290 Versoix, Switzerland}
\affiliation{Gravitational Wave Science Center (GWSC), Université de Genève, CH1211 Geneva, Switzerland}

\author[0000-0003-1474-1523]{Tassos~Fragos}
\affiliation{Département d’Astronomie, Université de Genève, Chemin Pegasi 51, CH-1290 Versoix, Switzerland}
\affiliation{Gravitational Wave Science Center (GWSC), Université de Genève, CH1211 Geneva, Switzerland}

\author[0000-0003-3684-964X]{Konstantinos~Kovlakas}
\affiliation{Institute of Space Sciences (ICE, CSIC), Campus UAB, Carrer de Magrans, 08193 Barcelona, Spain}
\affiliation{Institut d’Estudis Espacials de Catalunya (IEEC), Carrer Gran Capit\`a, 08034 Barcelona, Spain}

\author[0000-0001-9331-0400]{Matthias~U.~Kruckow}
\affiliation{Département d’Astronomie, Université de Genève, Chemin Pegasi 51, CH-1290 Versoix, Switzerland}
\affiliation{Gravitational Wave Science Center (GWSC), Université de Genève, CH1211 Geneva, Switzerland}

\author[0000-0003-4260-960X]{Devina~Misra}
\affiliation{Département d’Astronomie, Université de Genève, Chemin Pegasi 51, CH-1290 Versoix, Switzerland}
\affiliation{Institutt for Fysikk, Norwegian University of Science and Technology, Trondheim, Norway}

\author[0000-0002-0031-3029]{Zepei~Xing}
\affiliation{Département d’Astronomie, Université de Genève, Chemin Pegasi 51, CH-1290 Versoix, Switzerland}
\affiliation{Gravitational Wave Science Center (GWSC), Université de Genève, CH1211 Geneva, Switzerland}

\author[0000-0002-7464-498X]{Emmanouil~Zapartas}
\affiliation{Institute of Astrophysics, FORTH, N. Plastira 100,  Heraklion, 70013, Greece}
\affiliation{IAASARS, National Observatory of Athens, Vas. Pavlou and I. Metaxa, Penteli, 15236, Greece}

\begin{abstract}
We investigate Roche lobe overflow mass transfer (MT) in eccentric binary systems between stars and compact objects (COs), modeling the coupled evolution of both the star and the orbit due to eccentric MT (eMT) in a self-consistent framework.
We implement the analytic expressions for secular rates of change of the orbital semi-major axis and eccentricity, assuming a delta function MT at periapse, into the binary stellar evolution code \mesa{}.
Two scenarios are examined: (1) a simplified model isolating the effects of eMT on stellar and orbital evolution, and (2) realistic binary configurations that include angular momentum exchange (e.g., tides, mass loss, spin-orbit coupling, and gravitational wave radiation).
Unlike the ad hoc approach of instant circularization that is often employed, explicit modeling of eMT reveals a large fraction of binaries can remain eccentric post-MT. 
Even binaries which naturally circularize during eMT have different properties (donor mass and orbital size) compared to predictions from instant circularization, with some showing fundamentally different evolutionary outcomes (e.g., stable versus unstable MT).
We demonstrate that a binary's initial mass ratio and eccentricity are predictive of whether it will remain eccentric or circularize after eMT.
These findings underscore the importance of eMT in understanding CO-hosting binary populations, including X-ray binaries, gravitational wave sources, and other high-energy transients.
\end{abstract}


\section{Introduction}
\label{sec:intro}
Binary stellar interactions produce a wide array of high-energy astrophysical phenomena, including supernova explosions, X-ray binaries (XRBs), gravitational wave (GW) sources, and exotic stellar objects such as blue stragglers, hot subdwarfs and barium stars \citep[for reviews see; e.g.,][]{Bailyn1995,Heber2009ARAA,Langer2012,Tauris+2023pbse.book,Marchant_Bodensteiner2024ARAA,Wang2024}.
Primary among these interactions is \ac{MT} via \acf{RLO}, which has long been understood as integral in determining the evolution and ultimate fate of binary systems \citep{Kuiper1941ApJ,Paczynski1971ARAA,Eggleton1983ApJ,Webbink1984}.
The exchange of energy and angular momentum during \ac{RLO} \ac{MT} drives orbital evolution, producing a plethora of complex evolutionary sequences and phenomena \citep[e.g.,][]{Lubow1975,Kolb_Ritter1990AA}.
It is through this lenses that we interpret Low- and High-mass XRBs \citep[e.g.,][]{Bahramian_Degenaar.2023hxga.book,Kretschmar+2019NewAR,Fornasini+2023arXiv}, binary \ac{CO} mergers detected by the LIGO-Virgo-KAGRA collaboration \citep{Abbott+2023PhRvX..13a1048A}, transients such as supernovae with evidence for circumstellar interaction \cite[e.g.,][]{Moriya+2015A&A...584L...5M,Zhu+2024ApJ...970L..42Z}, wide binaries hosting quiescent \acp{CO} discovered with Gaia \citep[e.g.,][]{Andrews_Taggart_Foley2022arXiv,El-Badry2022MNRAS,El-Badry+2023MNRAS_BH1,El-Badry+2023MNRAS_BH2,GaiaCollab_Panuzzo+2024AA}, and gravitation wave sources which will be detectable by the Laser Interferometer Space Antenna (LISA) mission \citep[e.g.,][]{Lamberts+2019MNRAS.490.5888L,Tang+2024arXiv241102563T}.
Although extensively studied for over five decades, numerical models for binary evolution due to \ac{RLO} \ac{MT} have primarily considered circular orbits, neglecting eccentricity, a fundamental orbital property for many binary systems observed in nature.

Eccentric orbits are abundantly observed in both pre- and post-\ac{MT} binary systems; understanding how eccentric orbits may evolve through \ac{RLO} \ac{MT} is then paramount to understanding the nature of such systems.
During \acf{MS} evolution, many binaries are observed to have significantly eccentric orbits, which may provide both clues to their formation mechanisms and have consequences for their future evolution \citep{Mathieu1994,Geller2019,Munoz2020,D'Orazio2021,Hwang+2022MNRAS,Lubow2022,Bashi+2023MNRAS.522.1184B,Lai2023,Siwek2023,Hu+2024arXiv}. 
Just prior to the onset of \ac{MT}, ellipsoidal variables in the Large Magellanic Clouds retain significant eccentricity, in tension with standard tidal theory \citep[e.g.,][]{Nie+2017ApJ}.
High-mass X-ray binaries (\acsp{HMXB}) hosting \acp{CO} are observed in  eccentric orbits \citep[e.g.,][]{Raguzova_Popov2005AA,Neumann+2023AA,Fortin2023AA} and are expected to evolve into \ac{RLO} \ac{MT} when the donor leaves the \ac{MS} evolution while still eccentric \citep[][]{VanDenHeuvel.2019IAUS}.
Finally, post-\ac{MT} binaries hosting blue stragglers, barium, S-type stars and some post-AGB systems are observed to be in eccentric orbits \citep[e.g.,][]{Geller2009AJ,Milliman2014,Gosnell2015,Milliman2016,VanderSwaelmen+2017AA,Oomen+2018AA,Geller2021,Escorza2023,Nine+2024ApJ,Linck2024}, suggesting that a number of systems may gain or retain their eccentricity through \ac{RLO} \ac{MT}.

Although well motivated by observations, eccentric \ac{RLO} \ac{MT} is not trivially implemented into standard binary evolution calculations due to:
($i$) the commonly adopted Roche lobe model used to determine \ac{MT} conditions \citep{Eggleton1983ApJ,Ritter1988AA,Kolb_Ritter1990AA} may not apply in an eccentric orbit, as the resulting \ac{MT} interactions may require adoption of results from hydrodynamical simulations \citep[e.g. non-conservative \ac{MT} and re-accretion;][]{Lajoie_Sills2011ApJ,Saladino_Pols2019AA}, 
($ii$) the quantitative effects of \ac{RLO} \ac{MT} events on the long-term (secular) orbital evolution are challenging to calculate, depend on assumptions for \ac{MT} conditions along the orbit, and analytical expressions are needed for population studies \citep[e.g.,][and references therein]{Sepinsky+2007bApJ,Sepinsky2009ApJ,Dosopoulou_Kalogera2016ApJ,Hamers_Dosopoulou2019ApJ}.
Therefore, in detailed binary evolution simulations eccentricity is often not accounted for in \ac{MT} phases. 
Instead, binary orbits are either assumed to be circular at the \ac{ZAMS} of the binary system or, if later in the evolution \ac{MT} is encountered (at periapse) for eccentric binaries, then an ad hoc assumption of instantaneous circularization is made \citep[e.g.,][and others]{Hurley2002MNRAS,Naoz+2016ApJ,Ivanova+2005MNRAS,Eldridge+2017PASA...34...58E,Giacobbo+2018MNRAS.474.2959G,Sun2018,Breivik+2020ApJ...898...71B,Marchant2021AA,Sun2021,Rodriguez+2022ApJS,Fragos+2023ApJS,SimazBunzel+2023AA,Sun2023,Andrews+2024arXiv,Sun2024}.

In this work, we model the secular effects of \acf{eMT} on binary orbital evolution, using the 1D stellar evolution code \acl{mesa} (\mesa{}, version 11701) \citep{Paxton2011ApJS, Paxton2015ApJS, Paxton2013ApJS, Paxton2018ApJS, Paxton2019ApJS} along with the \mesa{} software development kit (version 20190503) \citep{richard_townsend_2020_3706650}. We also include other factors that impact mass and angular momentum evolution, such as tides, stellar winds, magnetic braking, and gravitational waves. 
Previous theoretical work involved deriving the proper secular effects of \ac{RLO} \ac{MT} in eccentric binary systems without the simultaneous evolution of the stellar companion \citep[e.g.,][]{Sepinsky+2007bApJ,Hamers_Dosopoulou2019ApJ}. 
We expand on this work, modeling the simultaneous evolution of both the star and orbit due to \ac{eMT}. 

In \autoref{sec:methods} we detail our choices for orbital evolution due to eccentric MT and implementation into \mesa{}.
In \autoref{sec:results:simple_case} we present a small parameter study with simplified models, to isolate the effects of eccentric MT.
In \autoref{sec:results:full_physics_models} we present our fiducial eccentric MT models with astrophysical initial conditions.
Finally, in \autoref{sec:discussion_conclusions} we discuss our findings and summarize with concluding remarks.

\section{Methods\label{sec:methods}}

\subsection{Treatment of Eccentric Mass Transfer}
\label{sec:methods:treatment_of_eMT}
We consider a binary with primary mass $M_1$, a secondary mass $M_2$, an orbital semi-major axis $a$, and an eccentricity $e$.
The total orbital angular momentum, $J_\mathrm{orb}$, is:
\begin{equation}
    J_\mathrm{orb} = M_1 M_2 \sqrt{ \frac{G a (1-e^2)}{M_1+M_2} },
    \label{eqn:Jorb}
\end{equation}
where $G$ is the gravitational constant.
The binary orbital period, $\Porb$, is related to the semi-major through Kepler's third law.
The time derivative of \autoref{eqn:Jorb} can be written as:
\begin{equation}
    \frac{\dot{J}_\mathrm{orb}}{J_\mathrm{orb}} = \frac{\dot{M}_1}{M_1} + \frac{\dot{M}_2}{M_2} - \frac{1}{2}\frac{\dot{M}_1 + \dot{M}_2}{M_1 + M_2} + \frac{1}{2}\frac{\dot{a}}{a} - \frac{e\dot{e}}{1-e^2}.
    \label{eqn:Jdot_div_J}
\end{equation}
In the classical case of a circular orbit with fully conservative \ac{RLO} \ac{MT} (where all transferred mass is accreted), the total mass is conserved with $\dot{M}_2 = - \dot{M}_1$, and the total orbital angular momentum is conserved, with $\dot{J}_\mathrm{orb}=0$.
Then, \autoref{eqn:Jdot_div_J} can be solved explicitly for the rate of change in the semi-major axis:
$\dot{a} = 2 a (q - 1)  \frac{\dot{M}_1}{M_1}$, 
where we define $q=M_1/M_2$ as the mass ratio.
As expected, the orbit shrinks when $q>1$ and expands when $q<1$.

In the case of an eccentric orbit, more complex methods are required to derive expressions for the secular rate of change of orbital elements $\dot{a}$ and $\dot{e}$. 
Such derivations have been carried out using variations of different formalisms \citep[e.g.,][and references therein]{Sepinsky+2007bApJ,Sepinsky2009ApJ,Sepinsky2010,Dosopoulou_Kalogera2016ApJ,Hamers_Dosopoulou2019ApJ}.
In this study, we incorporate the formalism from \citealt{Sepinsky2009ApJ} which is only analytical treatment currently available that can account for non-conservative \ac{MT} and can be applied to large-scale detailed binary evolution simulations.

We adopt Equations (18) and (19) from \citet{Sepinsky2009ApJ} which provide the secular rates of change for the semi-major axis $a$ and eccentricity $e$ as follows:
\begin{align}
    \left< \frac{da}{dt} \right>_\mathrm{sec} =  &\frac{a}{\pi} \frac{\dot M_0}{M_1} \frac{1}{ (1-e^2)^{1/2} } &\nonumber{}\\ 
    & \times \left[ e \frac{| \vec r_{\mathrm{A}_1,P} |}{a} + \gamma q e \frac{| \vec r_{\mathrm{A}_2} |}{a} \cos \Phi_P \right. &\nonumber{}\\ 
    &+ (\gamma q -1 ) (1-e^2) &\nonumber{}\\ 
    &\left. + (1 - \gamma) (\mu + \frac{1}{2}) (1-e^2) \frac{q}{q+1} \right],&
    \label{eqn:adot}
\end{align}
\begin{align}
    \left< \frac{de}{dt} \right>_\mathrm{sec} = &\frac{(1-e^2)^{1/2}}{2\pi} \frac{\dot M_0}{M_1} &\nonumber{}\\ 
    & \times \left[ \gamma q \frac{| \vec r_{\mathrm{A}_2} |}{a} \cos \Phi_P + \frac{| \vec r_{\mathrm{A}_1,P} |}{a} \right. &\nonumber{}\\ 
    &+ 2(\gamma q -1 ) (1-e) &\nonumber{}\\ 
    &\left. + 2(1 - \gamma) (\mu + \frac{1}{2}) (1-e) \frac{q}{q+1} \right],&
    \label{eqn:edot}
\end{align}
these equations can account for non-conservative \ac{MT}, where $\gamma=1-\beta$ is the fraction of mass transferred through \ac{RLO} that is accreted by the secondary ($\dot{M}_2= - \gamma \dot{M}_1$); $\mathrm{A}_1$ is the location where mass is lost from the donor at the $L_1$ point; $\mathrm{A}_2$ is the point on the accretor where mass is accreted; $\vec{r}_{\mathrm{A}_1}$ and $\vec{r}_{\mathrm{A}_2}$ are the position vectors of $\mathrm{A}_1$ and $\mathrm{A}_2$ relative to the centers of mass of the donor and accretor, respectively; $\Phi$ is the angle between the line connecting both component stars and the vector $\vec{r}_{\mathrm{A}_2}$; and subscript $P$ denotes a quantity is taken at periapse. 
Then, $\mu$ parametrizes the specific orbital angular momentum of material lost during non-conservative \ac{MT} which we take to be the specific orbital angular momentum of the accretor ($\mu=q$). 
We refer the reader to Figure 1 in \citealt{Sepinsky+2007bApJ} for definitions of all quantities listed here. 
Hereafter, Equations \ref{eqn:adot} and \ref{eqn:edot} will be denoted \adot{} and \edot{} respectively.

In this study, we model \ac{CO} accretors for which we assume $\vec{r}_{\mathrm{A}_2}/a \ll 1$ which implicitly accounts for $\Phi_\mathrm{P}$, following the simplified cases described in \citealt{Sepinsky+2007bApJ}.
Instead of assuming the star is exactly Roche-filling at all times during \ac{MT}, we explicitly calculate the \ac{MT} rate (Section \ref{sec:methods:calc_mdot_rate}).
The \ac{MT} efficiency $\gamma$ is assumed to be conservative up to the Eddington limit which is also explicitly calculated during evolution (\autoref{sec:methods:mesa_physics}).

The advantage of these secular evolution equations is that they are analytical while also accounting for non-conservative \ac{MT}, and they are derived under the following simplifying assumption: \ac{RLO} \ac{MT} occurs as a delta function at periapse with a $\ac{MT}$ rate $\dot{M}_1 = \dot{M}_0 \,\delta(\upsilon)$, where $\dot{M}_0$ is the instantaneous \ac{MT} rate, $\delta$ is the Dirac delta function, and $\upsilon$ is the true anomaly.
For a complete derivation and a list of all assumptions and their implications, we refer the reader to \citealt{Sepinsky+2007bApJ,Sepinsky2009ApJ}.
We note that while \citealt{Hamers_Dosopoulou2019ApJ} extend beyond the delta function \ac{MT} approximation with a phase dependent \ac{MT} rate, their formalism is limited to conservative \ac{RLO} \ac{MT} only.
Therefore, we use the equations by \citealt{Sepinsky2009ApJ}, as many binary systems are expected to undergo phases of non-conservative \ac{MT} which significantly impacts their evolution \citep[e.g.,][]{Marino+2017AA,Ziolkowski_Zdziarski.2018MNRAS}.

\subsection{Implementation in \mesa{}}
\label{sec:methods:orbital_evolution}
We incorporate the equations of secular orbital evolution due to eccentric \ac{MT} into \mesa{} by combining individual sinks (e.g. mass loss) and sources (e.g. mass accretion) of angular momentum transport. 
These are assumed to be decoupled, such that $\dot{J}_\mathrm{orb}$ can be determined by a linear combination of individual effects \citep[][]{Paxton2015ApJS}:
\begin{equation}
    \dot{J}_\mathrm{orb} = \dot{J}_\mathrm{gr} + \dot{J}_\mathrm{ml} + \dot{J}_\mathrm{mb} + \dot{J}_\mathrm{ls},
    \label{eqn:mesa_total_Jdot}
\end{equation}
where subscripts denote angular momentum changes due to gravitational wave radiation ($\dot{J}_\mathrm{gr}$), mass loss ($\dot{J}_\mathrm{ml}$), magnetic braking ($\dot{J}_\mathrm{mb}$), and spin-orbit coupling ($\dot{J}_\mathrm{ls}$).
Within $\dot{J}_\mathrm{ml}$, are contributions from circular non-conservative \ac{RLO} \ac{MT} and wind mass loss.
We replace the contribution from circular non-conservative \ac{RLO} \ac{MT}, with the contribution from eccentric \ac{MT} by reparameterizing \adot{} and \edot{} in terms of the orbital angular momentum via \autoref{eqn:Jdot_div_J}.

Similar to \autoref{eqn:mesa_total_Jdot}, we calculate the rate of change of orbital eccentricity, $\dot{e}_\mathrm{orb}$, as a linear combination of decoupled effects:
\begin{equation}
    \dot{e}_\mathrm{orb} = \dot{e}_\mathrm{gr} + \dot{e}_\mathrm{tidal} + \edot{},
    \label{eqn:mesa_total_edot}
\end{equation}
where $\dot{e}_\mathrm{gr}$ is from gravitational wave radiation, $\dot{e}_\mathrm{tidal}$ indicates tidal circularization, and $\edot$ is due to non-conservative eMT from \autoref{eqn:edot}. 
Models which circularize ($e=0$) have $\edot{}=0$ since the equations of \citealt{Sepinsky2009ApJ} are formally invalid in this regime.

\subsection{\mesa{} Binary and Stellar Evolution Physics}
\label{sec:methods:mesa_physics}
We use \mesa{} inlists developed for the state-of-the-art \ac{BPS} code \posydon{} version 2 (v2) \citep[][]{Andrews+2024arXiv}, which implements the results of detailed binary stellar evolution \mesa{} simulations in a framework for large-scale \ac{BPS}.
We leverage the \posydon{} software infrastructure to simulate and analyze \mesa{} simulations of H-rich stars with \ac{CO} companions at solar metallicity $Z = 0.0142$ \citep{Asplund2009ARAA}.
According to the adopted \mesa{} simulation setup, the binary is initialized in a tidally locked state with angular momentum loss occurring solely via stellar winds. 
Once the donor experiences \ac{RLO} (defined as $\log_{10}(\dot{M}/M_\odot\,\mathrm{yr}^{-1}) = -10$), we apply all sources of angular momentum transport including spin-orbit coupling, magnetic braking, and gravitational waves, in addition to mass loss.
This procedure was designed to match the properties of binaries determined to enter \ac{RLO} from population synthesis calculations in the \posydon{} framework \citep{Fragos+2023ApJS,Andrews+2024arXiv}.

We parameterized the tidal timescale developed from the linear theory \citep{Zahn.1977A&A....57..383Z,Hut1981A&A,Rasio+1996ApJ...470.1187R,Hurley2002MNRAS}, where the synchronization timescale of the entire model is then taken to be the mass-weighted average of synchronization timescales in each cell, taking the shortest between the dynamical and equilibrium tides.
We implement both spin-orbit coupling and (if the donor star has a mass $< 1.5 ~M_\odot$) magnetic braking.
For a complete description of all stellar and binary physics, we refer the reader to \citealt[][]{Fragos+2023ApJS,Andrews+2024arXiv}.

\acp{CO} are modeled as point masses (i.e., their properties other than mass are ignored) and are allowed to accrete up to their Eddington limit ($\dot{M}_\mathrm{Edd}$), above which material is assumed to be lost in an isotropic wind with the specific angular momentum of the accretor.
We define the Eddington limit via the relation: $\eta \dot{M}_\mathrm{Edd}c^2 = 4 \pi G M c / \kappa$, where $\eta$ is the efficiency of converting rest-mass energy to radiation defined as $\eta=\frac{G M}{R c^2}$, $c$ is the speed of light, $M$ and $R$ define the mass and accretion radius of the centra object, and we assume the Roseland mean opacity, $\kappa$, is dominated by Thompson scattering $\kappa=0.2(1+X)~\mathrm{cm^2~g^{-1}}$, where $X$ is the hydrogen abundance of the star. 
The \acp{CO} with masses below $2.5~M_\odot$ are assumed to be \acp{NS} with accretion radii of $12$ km, while \acp{CO} above $2.5~M_\odot$ are assumed to be \acp{BH} with accretion radii set by their innermost stable circular orbit \citep[see 4.2.3;][]{Fragos+2023ApJS}.

\subsubsection{Calculating the Mass Transfer Rate}
\label{sec:methods:calc_mdot_rate}
In the secular orbital evolution equations, we must determine the magnitude of the delta function \ac{MT} rate at periapse: $\dot{M}_0$.
We use the eccentric orbit-averaged \ac{MT} rate adapted from \cite{Ritter1988AA,Kolb_Ritter1990AA} as implemented in the \mesa{} \texttt{binary} module \citep{Paxton2015ApJS}.
The \ac{MT} rate is calculated at instantaneous points along the orbit assuming the standard Roche geometry \citep{Eggleton1983ApJ}:
\begin{equation}
    \mathcal{R}_L = D(\upsilon)\frac{0.49q^{2/3}}{0.6q^{2/3} + \ln(1+q^{1/3})},
    \label{eqn:roche_lobe_egg}
\end{equation}
where $D(\upsilon)=a(1-e^{2})/(1 + e\cos\upsilon)$ is the distance between the mass centers of the two objects as a function of the mean anomaly.
Although deviations from this Roche geometry are known to occur in eccentric orbits, \citealt{Sepinsky+2007aApJ} showed they are relatively small on the order of a few percent for a large range of $q$.
Additionally, other hydrodynamical effects may complicate the picture in determining an accurate \ac{MT} rate in eccentric orbits \citep[e.g.,][]{Lajoie_Sills2011ApJ,Davis_Siess_Deschamps2013AA,Saladino_Pols2019AA}, as well as detailed \ac{MT} stream trajectories \citep{Sepinsky2010}.
Due to these uncertainties, we leave more complex treatments of the MT to future work.
We discuss this more in \autoref{sec:discussion_conclusions}.

\subsection{\posydon{} Binary Population Synthesis Physics} \label{sec:methods:BPS_physics}
To obtain astrophysically motivated parameters for binaries hosting \acp{CO}, we perform an astrophysically motivated \ac{BPS} simulation with the state-of-the-art \posydon{} framework \citep{Fragos+2023ApJS,Andrews+2024arXiv}.
Our treatment of \ac{BPS} physics closely follows that of \citealt{Rocha+2024ApJ}, which showed agreement with many observed properties of the Galactic Be X-ray binary population.
However, we make one major change to their input physics, using the \ac{SN} remnant prescription of \citealt{Sukhbold+2016ApJ} which allows for the formation of low-mass \acp{BH}.
Additionally, we assume a burst (instead of constant) star formation history and evolve all binaries to $10\mathrm{\,Gyr}$ to sufficiently explore the range of binary configurations which experience \ac{eMT}.
In the following, we reiterate the other key physics assumed in our \posydon{} \ac{BPS} model, and refer a reader to \citealt{Rocha+2024ApJ} for a complete description.

Initial binary component masses and orbital configurations are drawn from observationally motivated distributions: primary masses are drawn from a Kroupa power-law \citep{Kroupa2001MNRAS} with $\alpha=2.3$ with $M_1/M_{\odot} \in [7-120]$, secondary masses are drawn uniformly $q\in[0,1]$ \cite{Sana2013AA} with a minimum of $1~M_\odot$, initial $\Porb/\mathrm{d} \in [1-10^{3.5}]$ following a flat-in-log distribution in the range [1-0.35] days following \citealt{Fragos+2023ApJS}, and eccentricities are all initially zero with stellar spins synchronized to the orbital period.

Common envelope evolution is modeled with the $\alpha_\mathrm{CE}-\lambda_\mathrm{CE}$ formalism \citep{Webbink1984,Livio_Soker1988ApJ}, where we set the fraction of orbital energy used to unbind the envelope $\alpha_\mathrm{CE}=1$ and the binding energy parameter $\lambda_\mathrm{CE}$ is calculated directly from stellar profiles from out binary \mesa{} simulations in \posydon{}.

Supernovae explosions are modeled using the remnant prescription of \citealt{Sukhbold+2016ApJ} for core-collapse SNe (CCSNe), and the remnant prescription of \citealt{Podsiadlowski2004ApJ} for electron-capture SNe (ECSNe) for He core masses between $1.4–2.5~M_\odot$.  
NSs are given natal kicks resulting from the \ac{SN} explosion drawn from a Maxwellian with a dispersion of $\sigma_\mathrm{CCSNe}= 265~\mathrm{km~s^{-1}}$ \citep{Hobbs2005MNRAS}, and a dispersion of $\sigma_\mathrm{ECSNe}= 20~\mathrm{km~s^{-1}}$ \citep{Giacobbo_Mapelli2019MNRAS}.
BHs receive natal kicks with the same dispersion $\sigma_\mathrm{CCSNe}$, scaled by the BH mass $M_\mathrm{BH}$ with $1.4~M_\odot / M_\mathrm{BH}$, following \cite{Fragos+2023ApJS}.

\subsubsection{Selecting Eccentric Binary Initial Conditions}
\label{sec:methods:popsyn_initial_conditions}
Since some fraction of binaries hosting \acp{CO} may naturally circularize from tidal evolution before initiating \ac{eMT}, we select a subset of binaries from our \posydon{} \ac{BPS} model \citep{Fragos+2023ApJS,Andrews+2024arXiv}.
We take binary parameters ($M_1$, $M_2$, $\Porb{}$, and $e$) at the onset of RLO (oRLO)\footnote{We choose a threshold between circular and eccentric binaries of $e=0.05$ since this more closely resembles the precision of observational measurements to distinguish such systems \citep[e.g.,][]{Fortin2023AA}.} for systems with $e>0.05$ (representing orbits that we consider as ``eccentric'') as initial conditions for our \ac{eMT} \mesa{} simulations to further investigate their evolution.
Within \posydon{}, oRLO in an eccentric orbit is determined when the star's radius is $95\%$ of the Roche lobe radius \autoref{eqn:roche_lobe_egg}.
We implicitly account for the evolutionary state of the donor at oRLO in \posydon{} (e.g. during post-\ac{MS} expansion) within the setup of our \mesa{} simulations (see \autoref{sec:methods:mesa_physics}).

In practice, the population of \ac{CO} and star binaries undergoing \ac{eMT} will depend strongly on the prior \ac{MT} physics and \ac{SN} model used in the \ac{BPS}.
Therefore, the comparison in this work between \ac{eMT} and the standard modeling choice adopted in \ac{BPS} (\autoref{sec:methods:inst_circ}) serves as a concrete analysis of the impact of \ac{eMT} modeling on \ac{CO}-hosting binaries.

\subsection{Comparison to Instant Circularization}\label{sec:methods:inst_circ}
Binaries which initiate \ac{MT} in eccentric orbits are almost always assumed to efficiently circularize due to strong tides that are expected in tight orbits or with evolved, radially extended stellar components \citep[e.g.,][]{Verbunt_Phinney.1995A&A...296..709V}.

Thus, the ad hoc assumption of \textit{instantaneous circularization} has been adopted throughout binary evolution simulations, where the initially eccentric orbit (with semi-major axis $a$ and eccentricity $e$) is instantly made circular with either: $(i)$ the new orbital separation, $a_\mathrm{c}$, being set equal to the periapse separation of the original orbit $a_\mathrm{c} = a(1-e)$, or $(ii)$ applying conservation of orbital angular momentum $a_\mathrm{c} = a({1 - e^2})$. 

To estimate the impact of our self-consistent treatment of \ac{eMT}, we run a set of \mesa{} simulations in both the original eccentric orbital configuration using Equations \ref{eqn:adot} and \ref{eqn:edot}, and the instantly circularized approximation $(i)$, using the periapse separation $a_\mathrm{c}=a(1-e)$.
We present these comparisons in \autoref{sec:results:full_physics_models}.

\begin{figure*}[ht]
    \centering
    \includegraphics[width=2\columnwidth]{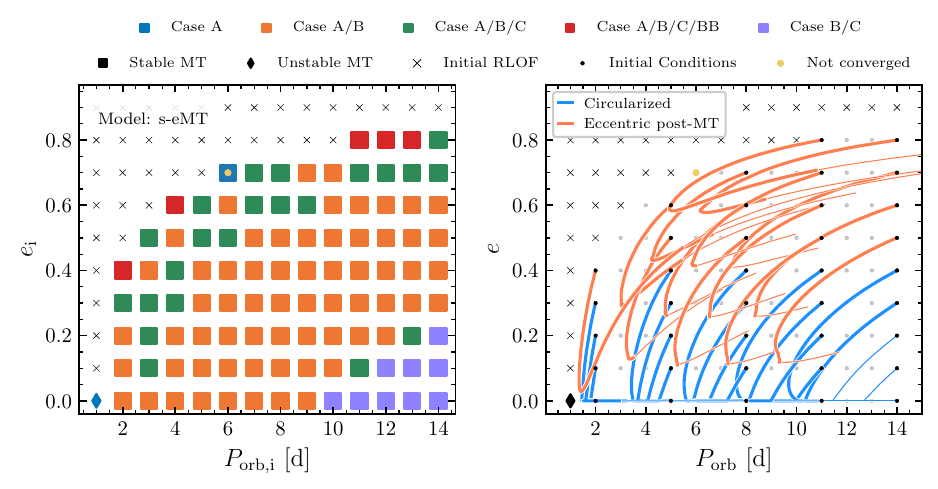}
    \caption{Results from our \ac{s-eMT} binary simulations with $20\,M_\odot$ donors and $10\,M_\odot$ BHs in various orbital configurations in the $\Porb{}{-}e$ plane, where all orbital evolution is due to the effects of non-conservative \ac{eMT} (\autoref{sec:results:simple_case}).
    The left panel shows the characteristic \ac{MT} history of our binary simulations (see legend and text), and the right panel shows the time evolution of a subset of models where thick (thin) lines correspond to the \ac{MS} (post-\ac{MS}) evolution of a given model (time evolution of gray models are not shown for clarity). 
    After \ac{MT}, systems with circularized orbits $(e = 0)$ are shown in blue, while those with non-zero eccentricities are shown in orange.
    }
    \label{fig:simple_models_ep_TF12}
\end{figure*}

\begin{figure*}
    \centering
    \includegraphics[width=\columnwidth]{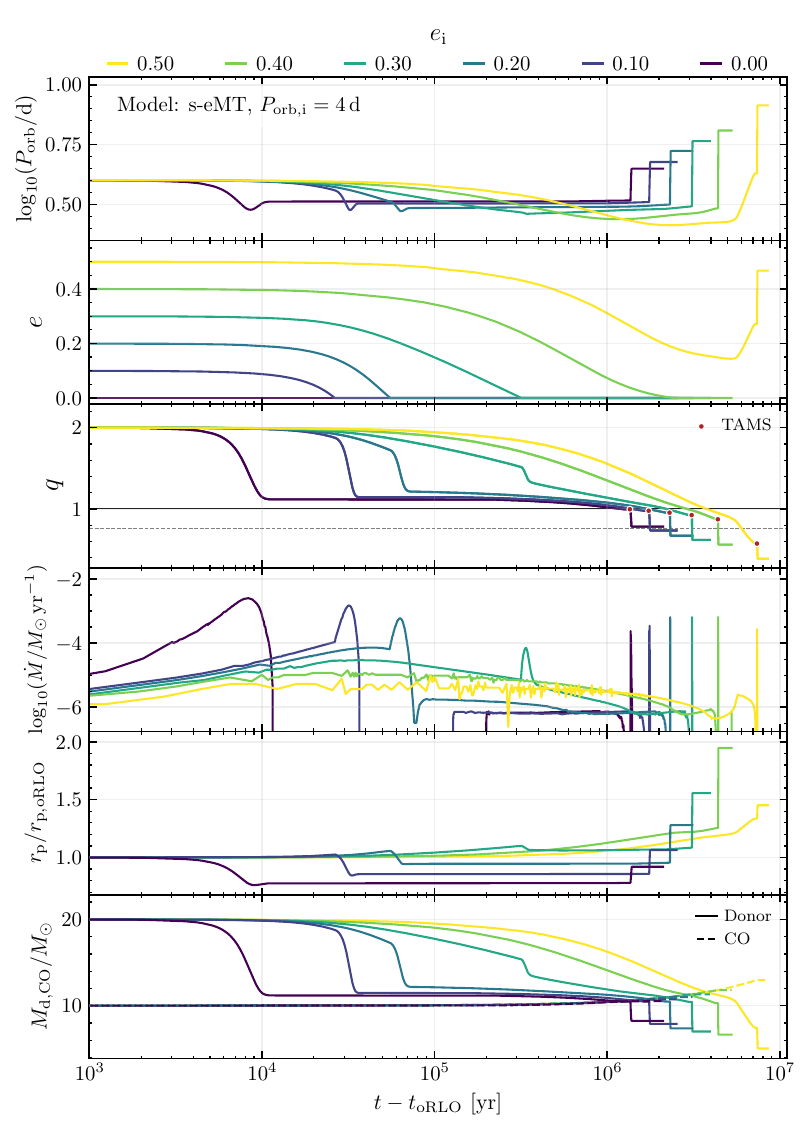}
    \includegraphics[width=\columnwidth]{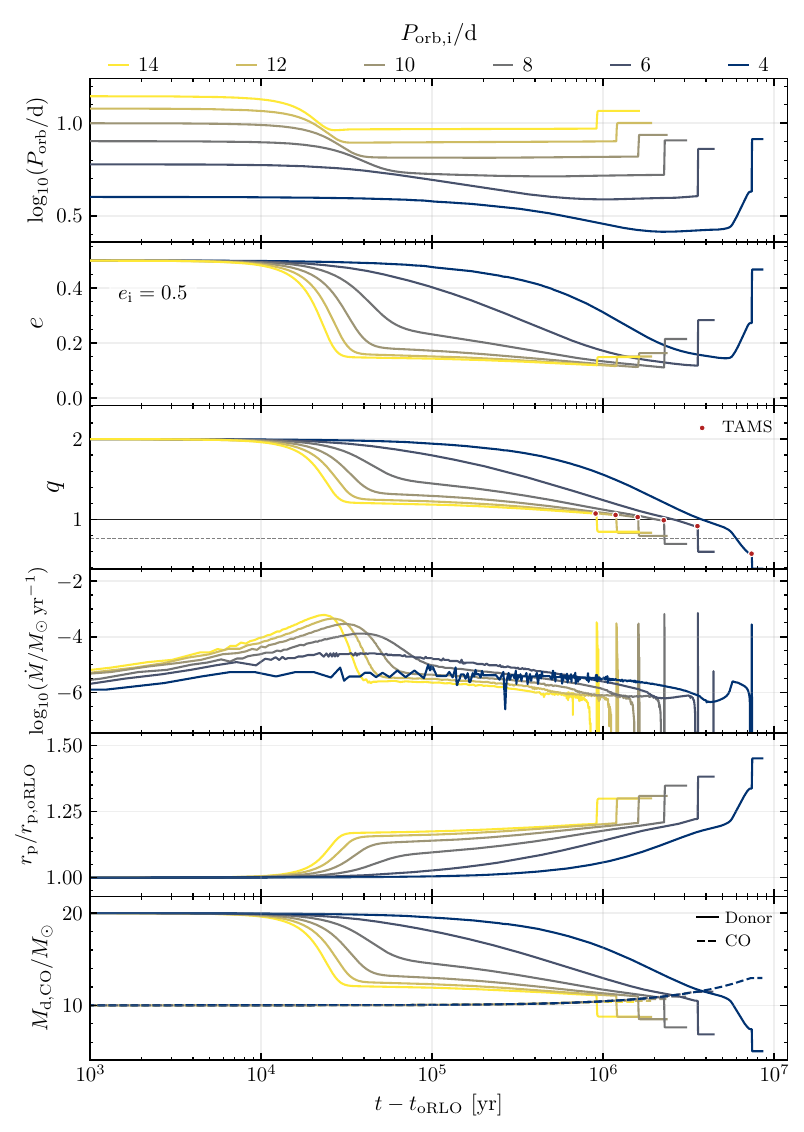}\hfill
    \caption{Time evolution from oRLO of the orbital period (first row), eccentricity (second row), mass ratio (third row), \ac{RLO} \ac{MT} rate (fourth row), the periapse separation normalized by the value at oRLO (fifth row), and the donor and \ac{CO} masses (last row), for a subset of our \ac{s-eMT} models with a $20\,M_\odot$ donor and $10\,M_\odot$ \ac{BH} shown in \autoref{fig:simple_models_ep_TF12}. 
    The colors in the left column indicate various initial eccentricities and various initial orbital periods in the right column.
    In the mass ratio ($q=M_1/M_2$) we have indicated with horizontal lines at $q=1$ (black) and $q=0.76$ (gray dashed) approximate transition points in the sign of \adot{} from \autoref{eqn:osep_q_crit} and \edot{} from \autoref{eqn:ecc_q_crit}.
    Red points in the $q$ panels denote when each model reaches \ac{TAMS} (thick to thin line transitions in \autoref{fig:simple_models_ep_TF12}), which occurs at different times due to the different $r_\mathrm{p,i}$ for each model.}
    \label{fig:simple_models_time_series}
\end{figure*}

\section{Parameter study with simplified eccentric mass transfer}
\label{sec:results:simple_case}
To isolate the effects of \ac{eMT}, we first run a suite of \mesa{} simulations with neglecting certain physical effects compared to our default evolutionary models \citep[described in \autoref{sec:methods:mesa_physics}; see also][]{Fragos+2023ApJS,Andrews+2024arXiv}.
These \acf{s-eMT} simulations neglect: 
    ($1$) donor rotation,
    ($2$) tidal coupling between the star and orbit,
    ($3$) tidal circularization, and
    ($4$) wind mass loss from the donor.
In practice, this \ac{s-eMT} setup is close to directly integrating \adot{} and \edot{}, but now the \ac{MT} rate is calculated self-consistently with a numerical stellar evolution model, unlike the procedure by \citealt{Sepinsky+2007bApJ,Sepinsky2009ApJ,Hamers_Dosopoulou2019ApJ}, where the \ac{MT} rate was kept constant without numerically solving a stellar structure model.

We consider binaries with a $10\,M_\odot$ \ac{BH} secondary and $20\,M_\odot$ \ac{MS} star primary at orbital periods $P_\mathrm{orb} \in [1{-}14]$ days and eccentricity $e \in [0{-}0.9]$.
Our choice of component masses and orbital periods are motivated by observed high-mass X-ray binaries in the Galaxy \citep{Neumann+2023AA,Fortin2023AA}, where the eccentricity is varied to facilitate the present parameter study.

In \autoref{fig:simple_models_ep_TF12}, we show the qualitative results of our simplified models in the $P_\mathrm{orb}{-}e$ plane, with the \ac{MT} history of each model indicated by colored symbols (left panel), and the time evolution of a subset of our binary models (right panel).
We first consider the \ac{MT} cases, using the standard definitions: Case A, Case B (Case BB), Case C, where each letter corresponds to oRLO during the \ac{MS}, before core helium depletion (from a stripped helium star), or in an evolved state post core helium exhaustion, respectively \citep{Iben1991ApJS}.
Systems that undergo multiple phases of \ac{RLO} are denoted by multiple letters separated by slashes (e.g. we denote Case A followed by Case B as Case A/B) as introduced in \cite{Fragos+2023ApJS}.

In the left panel of \autoref{fig:simple_models_ep_TF12}, all systems in our simplified model grid go through multiple phases of stable \ac{MT} (indicated by squares), except for the circular model at $P_\mathrm{orb,i}=1$ day which becomes dynamically unstable (shown by a diamond) while on the \ac{MS}.
Systems which have small initial periapse separations ($e_\mathrm{i} \gtrsim 0.2$ or $P_\mathrm{orb,i} \lesssim 10$ days) go through Case A MT, often followed by a second phase of Case B \ac{MT} initiated by post-MS expansion of the donor, and in systems with the highest eccentricity, Case C into Case BB \ac{MT} with an evolved helium-rich donor.
These \ac{MT} phases are consistent with other work investigating similar mass ratios in circular configurations \citep[e.g.,][]{Qin+2019ApJ,Marchant2021AA,Gallegos-Garcia+2021ApJ}, although the orbital evolution is altered due to our accounting for \ac{eMT}. 

In the right panel of \autoref{fig:simple_models_ep_TF12} we show how binaries with different initial conditions evolve during \ac{eMT}. 
We identify two types of evolutionary sequences based off the final orbital configuration of the simulations (prior to the core-collapse of the donor): (1) binaries which circularize due to \ac{RLO} \ac{MT} (blue), (2) binaries which remain eccentric post-\ac{MT} (orange).
Systems with initial eccentricities lower (higher) than $e \simeq 0.35$, end up circularizing (remain eccentric) by the end of the simulation. 
We note this circularization is due to the dynamics of \ac{eMT}, as tides are not included in these simplified runs. 
In our parameter study this bifurcation between circularized or eccentric orbits post-\ac{MT} is primarily a function of $e_\mathrm{i}$, although we expect this behavior may be sensitive to the mass ratio and $P_\mathrm{orb,i}$. 

It may also be seen with the time evolution in the $P_\mathrm{orb}{-}e$ plane that all models follow a similar qualitative behavior: MT initially drives the orbit to shrink (and circularize), but may be followed by a phase of orbital expansion and increasing eccentricity, appearing as a ``hook'' in the right panel of \autoref{fig:simple_models_ep_TF12}.
This behavior can be understood primarily through the evolution of the mass ratio, which has a significant impact on the sign of both \adot{} and \edot{}.
In the limit of conservative \ac{eMT}, \citealt{Sepinsky+2007bApJ} showed the transitional mass ratio, $q_\mathrm{crit}$, separating positive and negative rates of change for orbital parameters were found to be well approximated by their Equation (41) for \adot{}:
\begin{equation}
    q_\mathrm{crit} \simeq 1 - 0.4e + 0.18e^2,
    \label{eqn:osep_q_crit}
\end{equation}
and their Equation (42) for \edot{}:
\begin{equation}
    q_\mathrm{crit} \simeq 0.76 + 0.012e,
    \label{eqn:ecc_q_crit}
\end{equation}
where any $q$ above (below) these thresholds will have a positive (negative) rate of change.

To demonstrate this, in \autoref{fig:simple_models_time_series}, we show the time evolution of our \ac{s-eMT} simulations from oRLO of \Porb{}, $e$, $q$, \ac{MT} rate ($\dot{M}$), $r_\mathrm{p}$ normalized by the periapse separation at oRLO ($r_\mathrm{p,oRLO}$), and the component masses ($M_\mathrm{d}$ solid lines, $M_\mathrm{CO}$ dashed lines), for models with varying initial eccentricity (left panel) and initial orbital period (right panel). 
The line color emphasizes different initial conditions.
The switch from orbital shrinking and circularization to expansion and eccentricity pumping occurs near $q \simeq 1$ for \adot{} (solid black), and $q \simeq 0.76$ for \edot{} (gray dashed), both of which are achieved only after significant \ac{MT} occurs for this binary.

Focusing on systems with varying initial eccentricity and fixed $P_\mathrm{orb,i}$ (left column), binaries with higher $e_\mathrm{i}$ (smaller periapse distances) initiate MT earlier in the donor's \ac{MS} lifetime, which corresponds to progressively later times for reaching TAMS (red dots in $q$).
This behavior is also reflected in the MT cases in the left panel of \autoref{fig:simple_models_ep_TF12}, where models transition from Case A/B in the most eccentric binaries to Case B/C in the widest orbits (bottom right corner of the $\Porb{-}e$ plane).
Eccentric orbits also exhibit smaller peak \ac{MT} rates compared to their circular counterparts due to two effects: ($1$) the orbit averaged \ac{MT} rate is lower for more eccentric orbits (with fixed $r_\mathrm{p}$; see \autoref{eqn:roche_lobe_egg}), and ($2$) the orbital changes in $\Porb{}$ and $e$ result in an increase of the $r_\mathrm{p}$ separation, resulting in an effective widening of the orbit.
This second effect is unexpected compared to the classic result, which predicts orbital shrinking for $q>1$.
As discussed by \citealt{Sepinsky+2007bApJ}, this can be interpreted physically as a consequence of linear momentum exchange during the \ac{MT} at periapse.

This \textit{slower} Case A MT exhibited by more eccentric systems during \ac{eMT} (low peak $\dot{M}$ and longer \ac{MT} timescale) is more conservative than the systems which experience fast Case A MT. 
This allows the BHs to accrete more mass than predicted in circular systems (dashed lines in the last panels in \autoref{fig:simple_models_time_series}).
Binaries with the shortest initial periapse separations ($r_\mathrm{p,i}\lesssim 10~R_\odot$, or binaries which experience Case BB \ac{MT} in \autoref{fig:simple_models_ep_TF12}) can have BHs accrete up to $\sim5-9~M_\odot$, nearly doubling their initial mass.

This trend of earlier interaction leading to slower MT is also evident in the right column of \autoref{fig:simple_models_time_series}, where systems with initially short \Porb{} experience slow Case A MT, resulting in an anti-correlation between the initial \Porb{} and the final eccentricity of the binary.
Additionally, since the characteristic evolutionary timescale is governed by \ac{MT}, $\tau_\mathrm{MT} \propto \dot{M}^{-1}$, binaries with lower (higher) $\dot{M}$ evolve slower (faster). 
This effect can be observed comparing the models with initial $P_\mathrm{orb}$ values of $14$ and $4$ days, respectively, shown in the right column of \autoref{fig:simple_models_time_series}.

We highlight two key takeaways from this small parameter study: (1) not all binaries will circularize through \ac{eMT} and may even retain significantly eccentric post-MT, and (2) systems which do not circularize can end their evolution wider and more eccentric than their initial configuration.
We see strong correlations between the initial and final eccentricity and \Porb{} of our simplified models, but expect this behavior to be sensitive to the binary component masses, which we do not vary in our \ac{s-eMT} models.

\begin{figure*}[ht]
    \centering
    \includegraphics[width=2\columnwidth]{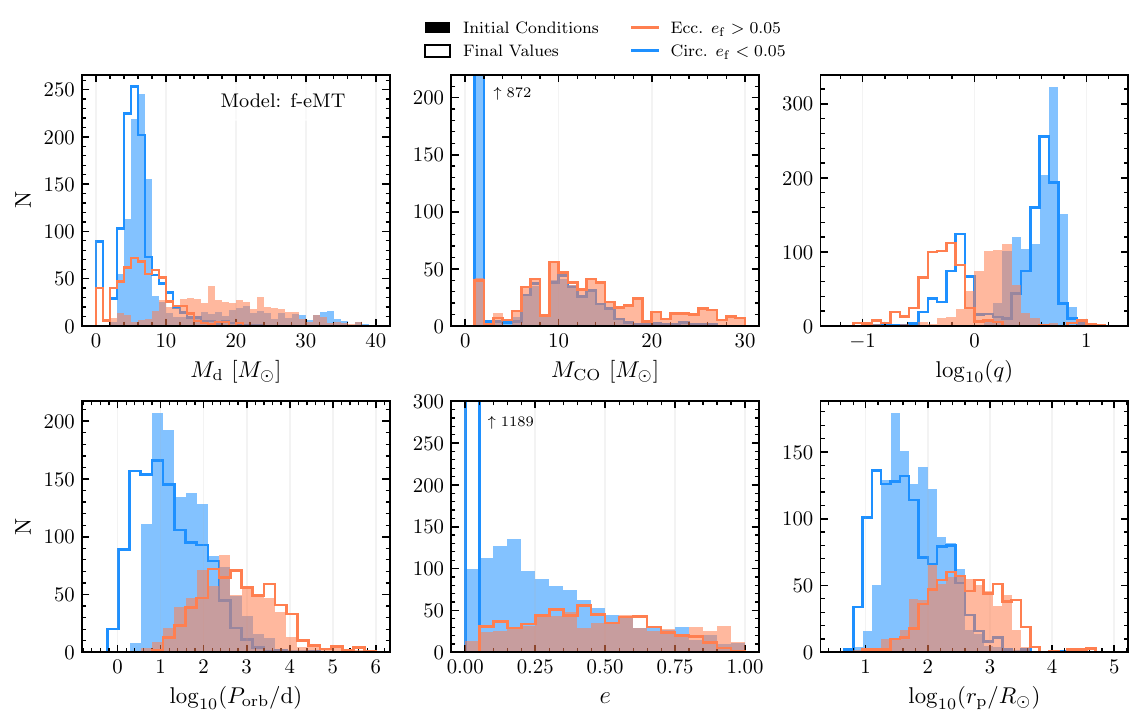}
    \caption{One-dimensional distributions of donor star mass (top left), CO mass (top middle), mass ratio between donor and CO mass (top right), orbital period (bottom left), eccentricity (bottom middle), and periapse distance (bottom right) for initial binary parameters (filled histograms) and final values (empty histograms) from our binary \ac{f-eMT} \mesa{} simulations, including secular orbital evolution due to \ac{eMT}, with self-consistent determination of the \ac{MT} rate, and additional sources of angular momentum loss (stellar winds, tides, magnetic braking, and gravitational wave radiation; see \autoref{sec:methods:mesa_physics}). 
    Binaries which remain eccentric post-MT are shown in orange and naturally circularized systems in blue, where initial conditions are drawn from an astrophysical \posydon{} BPS model (see Section \ref{sec:methods:popsyn_initial_conditions}).
    The end of a binary simulation can be triggered by the donor's imminent SN, WD formation, or the onset of unstable MT (not shown). 
    The eccentric post-MT binaries constitute $33\%$ of all our astrophysically sampled models. 
    } 
    \label{fig:1d_hist_IC_FV_ecc_circ_pops}
\end{figure*}

\section{Full eccentric mass transfer binary simulations}
\label{sec:results:full_physics_models}
With the characteristic effects of \ac{eMT} alone established in \autoref{sec:results:simple_case}, we now run a suite of binary simulations taking into account an expanded set of physical effects, as considered in the \posydon{} models: stellar rotation, tidal coupling between the stellar spin and orbital motion, mass and angular momentum loss via stellar winds, as well as angular momentum loss from magnetic braking and gravitational wave radiation.

We consider a population of two hydrogen-rich stars with relatively massive primaries with typical initial conditions (see Section \ref{sec:results:initial_conditions_from_popsynth}) which evolves through to the formation of the population's first \acp{CO} (\acp{NS} or \acp{BH}). 
Following \ac{CO} formation, most surviving binaries have eccentric orbits due to both mass loss and potential natal kicks imparted to the \acp{CO} at core collapse.
We follow the evolution of systems through their detached phase \citep[details in,][]{Fragos+2023ApJS,Andrews+2024arXiv} until \ac{RLO} is initiated at periapse.
At this point we simulate the population in two different ways for comparison: ($i$) through \ac{eMT} but this time accounting for the expanded set of physics as aforementioned, and we call these models \acf{f-eMT} models, and ($ii$) with the standard BPS ad hoc assumption of instant circularization.
This allows us to assess the impact of \ac{f-eMT} on a realistic, typical population of binaries harboring one compact object and a non-degenerate companion and compare results with the instant circularization assumption that is often made. 

For our analysis, we separate \ac{f-eMT} binaries into two broad outcomes: the population that remains eccentric post-MT in \autoref{sec:results:ecc_post-MT} and those that naturally circularize in \autoref{sec:results:natty}. 
The former outcome is clearly not produced in \ac{BPS} to date, as systems are forced into circularization. 
The latter outcome produces circularized systems through\ac{eMT}; we directly compare these to the circular population from the instant-circularization simulations (\autoref{sec:methods:inst_circ}).

\subsection{Eccentric Binary Population Properties}
\label{sec:results:initial_conditions_from_popsynth}

From our astrophysical \posydon{} \ac{BPS} simulation (\autoref{sec:methods:BPS_physics}), we select models which are instantaneously circularized in \posydon{} with a non-degenerate star and \ac{CO} companion in eccentric orbits during their detached phase post-SN (Section \ref{sec:methods:popsyn_initial_conditions}).
From our initial binary population of $4\times10^6$ binaries, $16\%$ survive interactions from the primary's \ac{SN} to form binaries containing a \ac{CO} and a star, hosting either \acp{NS} or \acp{BH}.
Of all the \ac{CO}-hosting binaries, we find $81\%$ enter \ac{RLO} at some point in their evolution. 
About $3/4$ ($72\%$) of these \ac{RLO} systems have residual eccentricity at oRLO ($e>0.05$, Section \ref{sec:methods:popsyn_initial_conditions}), which may be considered nominally as eccentric orbits (we take a total of 1791 pair models for direct comparison between \ac{f-eMT} and instant-circularization).
Thus, a majority of detached binaries which subsequently undergo \ac{MT} between a CO and a star in our fiducial \ac{BPS} model would be subjected to the ad hoc treatment of instantaneous circularization at oRLO, neglecting eMT outcomes.

In \autoref{fig:astro_pop_to_MESA} we show one-dimensional distributions of the binary population which undergo eccentric \ac{RLO} from \posydon{} (black), and the initial $2000$ systems we randomly sample for our \mesa{} models with \ac{f-eMT} (orange).
Our binary models cover a wide range of donor masses ($M_\mathrm{d}/M_\odot \in [1{-}50]$), \ac{CO} masses with NSs and BHs ($M_\mathrm{CO}/M_\odot \in [1.1{-}34]$), and orbital configurations ($\Porb{}/\mathrm{d} \in [2{-}10^5]$, $e \in [0.05, 0.99]$).
The top right panel shows the resulting range of mass ratios with two prominent peaks at $q \simeq 1-2$ from predominantly \ac{BH}-hosting binaries, and $q \simeq 3-8$ with NS-hosting binaries.
The most common configuration for a \ac{CO} and star binary to initiate \ac{eMT} has a \Porb{} of $10$ days, a modest eccentricity $e \simeq 0.15$, with a $1.2\,M_\odot$ NS and $7\,M_\odot$ donor.
The overlap of our samples and the underlying astrophysical population show good agreement within the Poisson errors (vertical bars), verifying that our subset of \mesa{} models is representative of the astrophysical \ac{BPS}.

\subsection{Binaries that Remain Eccentric Post-MT}
\label{sec:results:ecc_post-MT}
We first focus on binaries which remain eccentric post-MT with $e_\mathrm{f}>0.05$, as such systems cannot form under the assumptions of instantaneous circularization. 
These systems demonstrate that tides and/or \ac{eMT} itself do not efficiently circularize all binaries.
In \autoref{fig:1d_hist_IC_FV_ecc_circ_pops}, we show the initial conditions (filled histogram) and final properties (empty histograms) of this eccentric post-MT population in orange, comprising $33\%$ of the astrophysical population. 
In all these binaries, the \ac{MT} remains stable, with most of them ($95\%$) initiating \ac{MT} during the donor's post-MS evolution (Case B MT). 

\begin{figure}[t]
    \centering
    \includegraphics[width=1\linewidth]{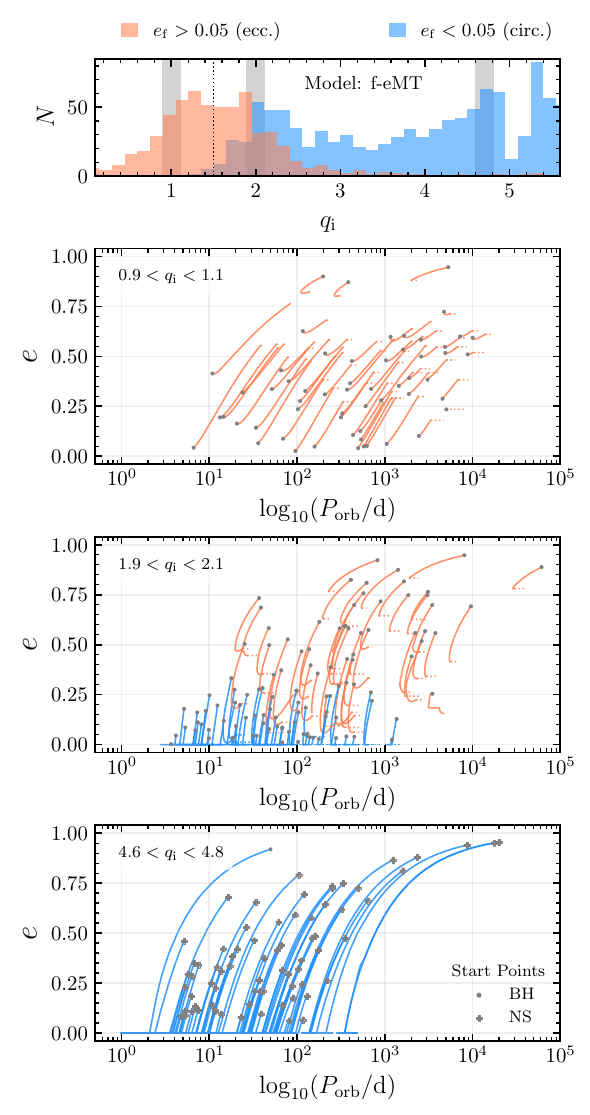}
    \caption{Orbital evolution in the $P_\mathrm{orb}{-}e$ plane for binaries (bottom three panels, for three different initial mass ratio ranges) selected by their initial mass ratio $q_\mathrm{i}$ (top panel), where binaries which circularize or remain eccentric post-MT are shown in blue and orange respectively (as in \autoref{fig:1d_hist_IC_FV_ecc_circ_pops}).
    Solid lines indicate active RLO \ac{MT} while dotted lines indicate post-MT detached evolution. Systems containing a BH (NS) are marked with a grey dot (plus) at the beginning of the evolution trajectory.
    The cause of the `hook’ evolution is similar to that in our simplified models (see \autoref{fig:simple_models_ep_TF12}, occurring near $q=1$), but with notable differences due to the inclusion of additional physical effects (e.g., wind mass loss).
    } \label{fig:Porb-e-split_q}
\end{figure}

This population has low $q_\mathrm{i}$ (upper right panel, orange filled histogram), and is therefore dominated by BH-hosting binaries with high mass donors ($M_\mathrm{d}>10\,M_\odot$).
Binaries with low mass ratios require less \ac{MT} to reach the regime where {\edot{}} changes sign from negative to positive ($q \simeq 0.76$, \autoref{eqn:ecc_q_crit}), thereby entering a regime  where their eccentricity increases rather than reduces, avoiding circularization.
Since \ac{MT} is initiated post-MS, it proceeds rapidly (on a thermal timescale) leading to non-conservative MT (inefficient accretion), explaining the relatively small changes in CO masses (top center panel of \autoref{fig:1d_hist_IC_FV_ecc_circ_pops}).
This MT history produces stripped, helium-rich donors with two peaks in the final mass distribution: $M_\mathrm{d} \lesssim 1\,M_\odot$ forming \acp{WD}, and a wider peak around $M_\mathrm{d} \simeq 6\,M_\odot$, which are massive enough to experience core collapse.

The orbital configurations of the eccentric post-MT population are characterized by wider orbits ($P_\mathrm{orb,i}/\mathrm{d} \simeq 10-10^4$, and $\log_{10}(r_\mathrm{p}/R_\odot) > 1.5$) with a nearly flat initial eccentricity distribution.
At the population level, there is relatively little change between the initial and final orbital parameter distributions, although individual binaries can experience significant evolution in both $P_\mathrm{orb}$ and $e$ over time.

To demonstrate this, we show in \autoref{fig:Porb-e-split_q} the correlated evolution in the $P_\mathrm{orb}{-}e$ plane (bottom three panels), selected by their initial mass ratio (corresponding grey bins in the top panel), where orange and blue lines correspond to the eccentric and circularized populations, respectively (as in \autoref{fig:1d_hist_IC_FV_ecc_circ_pops}) with starting points of evolution for BH- and NS- hosting binaries marked by circles and pluses, respectively.
Similar to the evolutionary sequences identified for our simplified \ac{s-eMT} models in \autoref{sec:results:simple_case}, the \ac{f-eMT} models exhibit phases of both orbital expansion/shrinking and eccentric circularization/pumping primarily understood through the evolution in $q$.

We find that binaries with low $q_\mathrm{i} \simeq 1$ (predominantly BH-hosting), remain eccentric throughout their evolution, even post-MT. 
The reason is that at these mass ratios the values of \adot{} and \edot{} promptly turn positive with very little \ac{MT}, leading to movement toward the upper right corner of the $P_\mathrm{orb}{-}e$ plane.
This correlated movement causes BH-binaries to end their evolution at larger orbital periods ($P_\mathrm{orb} \simeq 10^2 - 10^4$ days) with more eccentric ($e\simeq0.2-0.8$) orbits.
Binaries with larger $q_\mathrm{i} \gtrsim 2-3$ (similar to the \ac{s-eMT} models in \autoref{fig:simple_models_ep_TF12}) initially have negative \adot{} and \edot{}, with only highly eccentric binaries escaping complete circularization.
We note that overall, binaries which remain eccentric post-MT but have large $q_\mathrm{i}$ end their evolution in both tighter and wider orbits, but generally experience eccentricity dampening, where the end of MT influences $e_\mathrm{f}$ (see binaries with ``hook'' evolution in the center $\Porb{}{-}e$ panel of \autoref{fig:Porb-e-split_q}).

\begin{figure}[t]
    \centering
    \includegraphics[width=1\linewidth]{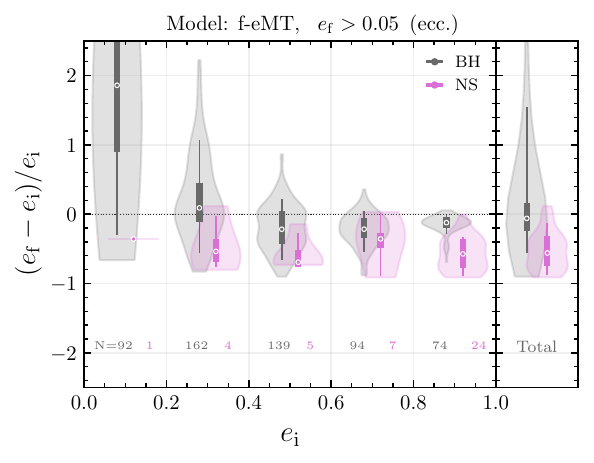}
    \caption{For the eccentric post-MT population (orange data in \autoref{fig:1d_hist_IC_FV_ecc_circ_pops}): distributions of the fractional change in eccentricity as a function of $e_\mathrm{i}$, for binaries hosting \acp{BH} (gray lines and markers) and \acp{NS} (pink lines and markers), with models binned in intervals of $e_\mathrm{i}=0.2$, with a small offset in each bin for visual clarity. 
    The full marginal distribution is shown on the right. 
    The median, $68\%$, and $90\%$ percentiles are shown as circles, thick lines, and thin lines, respectively, while the violin plot shows a kernel density estimate from all points in a bin. 
    Sample counts ($\mathrm{N}$) for BHs and NSs respectively are provided below each violin.
   } 
    \label{fig:delta_eccentricity}
\end{figure}

To understand how models that remain eccentric post-MT ($e_\mathrm{f} > 0.05$, shown as orange data in Figures \ref{fig:1d_hist_IC_FV_ecc_circ_pops} and \ref{fig:Porb-e-split_q}) increase or decrease their eccentricity, in \autoref{fig:delta_eccentricity} we present the distribution of fractional changes in $e$ from initial to final values through \ac{eMT}, $(e_\mathrm{f}-e_\mathrm{i})/e_\mathrm{i}$, for binaries hosting \acp{BH} (gray) and \acp{NS} (purple). 
The fractional change here is meant to quantify the degree to which a binary system's eccentricity falls above or below its initial value in our \ac{f-eMT} models.
%
%

Binned by their initial eccentricity, we identify a correlation in BH-hosting systems where low $e_\mathrm{i}$ trends towards greater positive fractional changes (i.e., trends towards $e_\mathrm{f} > e_\mathrm{i}$). 
Moving towards greater $e_\mathrm{i}$, we see this fractional change decreases towards zero and slightly negative at the greatest initial eccentricities ($e_\mathrm{i} \gtrsim 0.8$). Thus, systems with greater $e_\mathrm{i}$ show relatively minor changes in their eccentricity through \ac{eMT}.
The variance in the fractional change in $e$ decreases significantly with increasing $e_\mathrm{i}$, with $90\%$ of binaries having fractional changes $\lesssim \pm 25\%$ for $e_\mathrm{i} > 0.8$.
NS-hosting binaries trend toward smaller eccentricities post-MT (i.e., $e_\mathrm{f} < e_\mathrm{i}$), which may be seen in the marginal distributions showing the overall trends for BH- and NS-hosting binaries on the right side of \autoref{fig:delta_eccentricity}. 
This is expected from the larger $q_\mathrm{i}$ of NS-hosting systems, but a few eccentric pumping outcomes (i.e., $e_\mathrm{f} > e_\mathrm{i}$) occur at moderate eccentricities for this subpopulation.
NS-hosting binaries which remain eccentric post-MT are more prevalent at higher $e_\mathrm{i} \gtrsim 0.6$, as they can undergo more \ac{MT} before circularization ($q$ thresholds from \autoref{eqn:ecc_q_crit}).

\begin{figure}[t]
    \centering
    \includegraphics[width=1\linewidth]{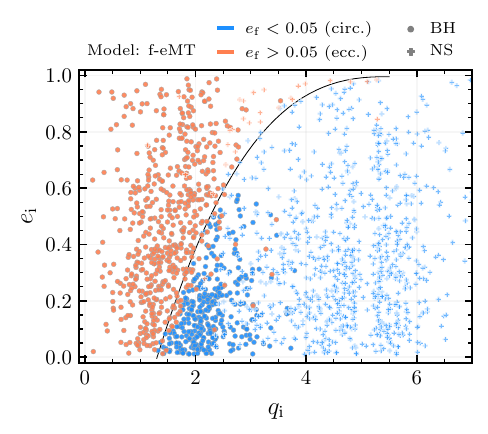}
    \caption{Initial conditions for our \ac{f-eMT} models in initial $q-e$ plane, color-coded by models that either naturally circularize (blue) or remain eccentric (orange) post-MT as described in \autoref{fig:1d_hist_IC_FV_ecc_circ_pops}. 
    Markers indicate binaries hosting BHs (circles) and NSs (pluses). 
    A bifurcation eccentricity, $e_\mathrm{crit}$, which distinguishes binaries that circularize from those that remain eccentric post-MT, is indicated by the black line. 
    A corresponding fitting function is provided in \autoref{eqn:q_e_polyfit}.
    For BH-hosting binaries, $64\%$ end up remaining eccentric post-MT, while $36\%$ naturally circularize.}
    \label{fig:bifurcation_q_e}
\end{figure}

Motivated by the trends identified in the $P_\mathrm{orb}{-}e$ plane (\autoref{fig:Porb-e-split_q}), we investigate the role of the initial mass ratio, $q_\mathrm{i}$, on determining outcomes of \ac{f-eMT} simulations.
We show in \autoref{fig:bifurcation_q_e} our \ac{eMT} models in the $q_\mathrm{i}{-}e_\mathrm{i}$ plane, where we find a clear bifurcation between the eccentric and circularized post-MT populations.
We perform a polynomial fit to this bifurcation, characterized as an eccentricity, $e_\mathrm{crit}$, that is a function of the mass ratio $q$:
\begin{align}
    e_\mathrm{crit} \simeq& -0.210 - 0.729\,q -0.1444\,q^2 - 0.0093\,q^3
    \label{eqn:q_e_polyfit}
\end{align}
derived from simulations in the range $q_\mathrm{i} \in [0,5.5]$.
Separating the \ac{CO} populations, $64\%$ of the BH-hosting binaries remain eccentric post-MT compared to only $5\%$ for NS-hosting binaries.

The bifurcation in $q_\mathrm{i}{-}e_\mathrm{i}$ is evidence that the evolution of our \ac{f-eMT} models is primarily impacted by the secular orbital evolution from \ac{eMT} (from \autoref{eqn:adot} and \autoref{eqn:edot}) compared to the other sources of angular momentum exchange (stellar winds, tides, magnetic breaking, and gravitational waves).
Applying \autoref{eqn:q_e_polyfit} to our \ac{s-eMT} models (\autoref{sec:results:simple_case}) with $q_\mathrm{i}=2$, the predicted bifurcation eccentricity $e_\mathrm{crit} \simeq 0.4$ agrees reasonably well with the results in Figures \ref{fig:simple_models_ep_TF12} and \ref{fig:simple_models_time_series}.
The shape of \autoref{eqn:q_e_polyfit} can be understood to first order, as determined by the competing timescales of changes in eccentricity ($\tau_e \simeq e/\dot{e}$) and mass ratio ($\tau_\mathrm{MT} \simeq M/\dot{M}$, which can be seen in the prefactor of \autoref{eqn:edot}).
Binaries with higher $e_\mathrm{i}$ can undergo more \ac{MT} before complete circularization, reaching low enough $q$ values to exhibit eccentric pumping (the ``hook'' evolution in the $P_\mathrm{orb}{-}e$ plane from \autoref{fig:Porb-e-split_q}).
Although, analytically deriving this relation is non-trivial due to the inclusion of non-conservative \ac{MT} and other effects simultaneously evolved within \mesa{}.
The few outlier \ac{BH} models that remain eccentric post-MT but lie in the high $q_\mathrm{i}\simeq3$ regime (expected to circularize), remain eccentric due to extremely strong stellar winds from high mass donors ($\gtrsim 20\,M_\odot$), which significantly alter their orbit and mass ratio.

\begin{figure*}[t]
    \centering
    \includegraphics[width=2\columnwidth]{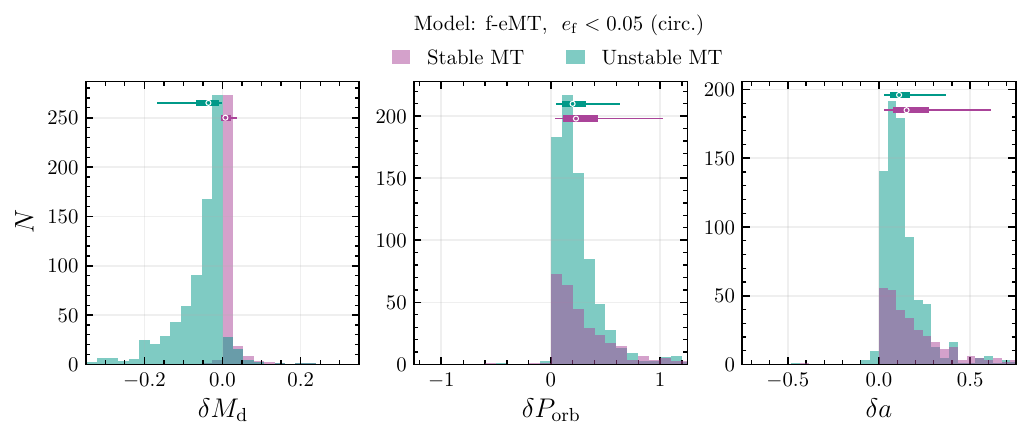}
    \caption{For models which naturally circularize in our \ac{f-eMT} simulations (blue in \autoref{fig:1d_hist_IC_FV_ecc_circ_pops}): fractional change (\autoref{eqn:frac_diff_circularized}) in donor mass (left panel), orbital period (center panel), and semi-major axis (right panel) relative to the instantaneously circularized treatment at the end of each simulations.
    Positive (negative) values indicate increases (decreases) in a given parameter when using \ac{f-eMT} compared to instant circularization. 
    Colors indicate simulations where both models undergo stable (purple) or unstable \ac{MT} (green), where the distribution median, $68\%$, and $90\%$ percentiles are shown near the top of each histogram (as described in \autoref{fig:delta_eccentricity}).
    } 
    \label{fig:circularized_1d_fracdiffs}
\end{figure*}

\subsection{Binaries that Naturally Circularize}
\label{sec:results:natty}
We now investigate the subset of binaries that naturally circularize due to \ac{eMT} (blue distribution in \autoref{fig:1d_hist_IC_FV_ecc_circ_pops}), which make up $66\%$ of all our binary simulations sampled with astrophysical initial conditions.
These systems are those that can be directly compared to the circular systems produced in the standard \ac{BPS} population through ad hoc instant-circularization. 
These naturally circularized binaries from the \ac{f-eMT} models generally have large initial mass ratios ($q_\mathrm{i} \gtrsim 4$, which are predominantly NS-hosting binaries), lower $e_\mathrm{i}$ (BH-hosting binaries with small natal kicks), or a combination of both features.
This is evident in the $\Porb{}{-}e$ evolution in \autoref{fig:Porb-e-split_q} and in the $q_\mathrm{i}{-}e_\mathrm{i}$ plane from \autoref{fig:bifurcation_q_e}.
The effects which cause binaries to circularize are described in detail by the evolution in the $P_\mathrm{orb}{-}e$ plane in \autoref{fig:Porb-e-split_q} and the bifurcation in the $q_\mathrm{i}-e_\mathrm{i}$ plane in \autoref{fig:bifurcation_q_e}.

First, we consider the MT histories of the models (as in the left panel of \autoref{fig:simple_models_ep_TF12}), which are summarized in \autoref{table:MT_cases_circularized}.
We find $96\%$ of models agree in their MT history (both evolving through stable or entering dynamically unstable MT) with the remaining $4\%$ having completely divergent evolution.
For pair models which agree, we also find that MT cases between models are consistent. 
In other words, the type of MT (e.g., Case A, Case B, and Case C) is the same when running the instantly circularized \mesa{} binary model versus \ac{f-eMT}.
Finally, for the pair models which diverge, our \ac{f-eMT} treatment results in more stable MT ($\sim 72\%$) compared to the instant circularization treatment leading to unstable MT (i.e., common envelope evolution; $\sim 21\%$).
This preference for stability is due to the $r_\mathrm{p}$ evolution seen in \autoref{fig:simple_models_time_series}, which effectively widens the binary while residual eccentricity remains in the orbit.

We also consider the differences in final properties of our naturally circularized models and their corresponding models which were instantly circularized (\autoref{sec:methods:inst_circ}).
Our binary \mesa{} simulations are stopped based on variety of criteria related to the formation of a \ac{WD}, the imminent core collapse of the primary, and the onset of dynamically unstable MT \citep[for details, see][]{Fragos+2023ApJS,Andrews+2024arXiv}.
Thus, we compare the final outcomes of a quantity $\mathbf{x}$ between our \ac{f-eMT} simulations ($\mathbf{x}_\mathrm{eMT}$) and their equivalent instantly circularized \mesa{} models ($\mathbf{x}_\mathrm{instant\,circ}$) by computing the fractional changes between the two cases which we denote with $\delta$:
\begin{equation}
    \delta \mathbf{x} = \frac{\mathbf{x}_\mathrm{eMT} - \mathbf{x}_\mathrm{instant\,circ}}{\mathbf{x}_\mathrm{instant\,circ}},
    \label{eqn:frac_diff_circularized}
\end{equation}
where we compute the differences relative to the instantly circularized case, as this is the standard assumption in binary stellar evolution.
For clarity, we only compute differences between models with matching MT histories (both stable or unstable MT, see \autoref{table:MT_cases_circularized}).

In \autoref{fig:circularized_1d_fracdiffs}, we present one-dimensional histograms for the fractional differences in final donor mass $\delta M_\mathrm{d}$, orbital period $\delta P_\mathrm{orb}$, and orbital separation $\delta a$, separated by stable or unstable MT.
The binaries undergoing stable MT may go on to experience the donor's core collapse or \ac{WD} formation, while binaries undergoing unstable MT enter common envelope evolution.

Our naturally circularized \ac{f-eMT} simulations predict similar donor masses to the instantly circularized case, with $90\%$ of models having $\delta M_\mathrm{d} \lesssim \pm 20\%$, with medians at about $\lesssim \pm 5\%$. 
The stable MT binaries have smaller $\delta M_\mathrm{d}$ and are positively skewed compared to unstable MT models which have larger absolute $\delta M_\mathrm{d}$ skewing negative.
Additionally, \ac{f-eMT} simulations overwhelmingly predict $\delta P_\mathrm{orb} > 0$, for both stable and unstable MT distributions, with much larger fractional changes ($90\%$ from $\delta P_\mathrm{orb} \simeq 50-100\%$ and $\delta a \simeq 25-55\%$).
Therefore, even binaries which naturally circularize during \ac{eMT} have different parameters compared to the ad hoc instant circularization approach for both unstable and stable MT systems.

\begin{figure*}[ht]
    \centering
    \includegraphics[width=2\columnwidth]{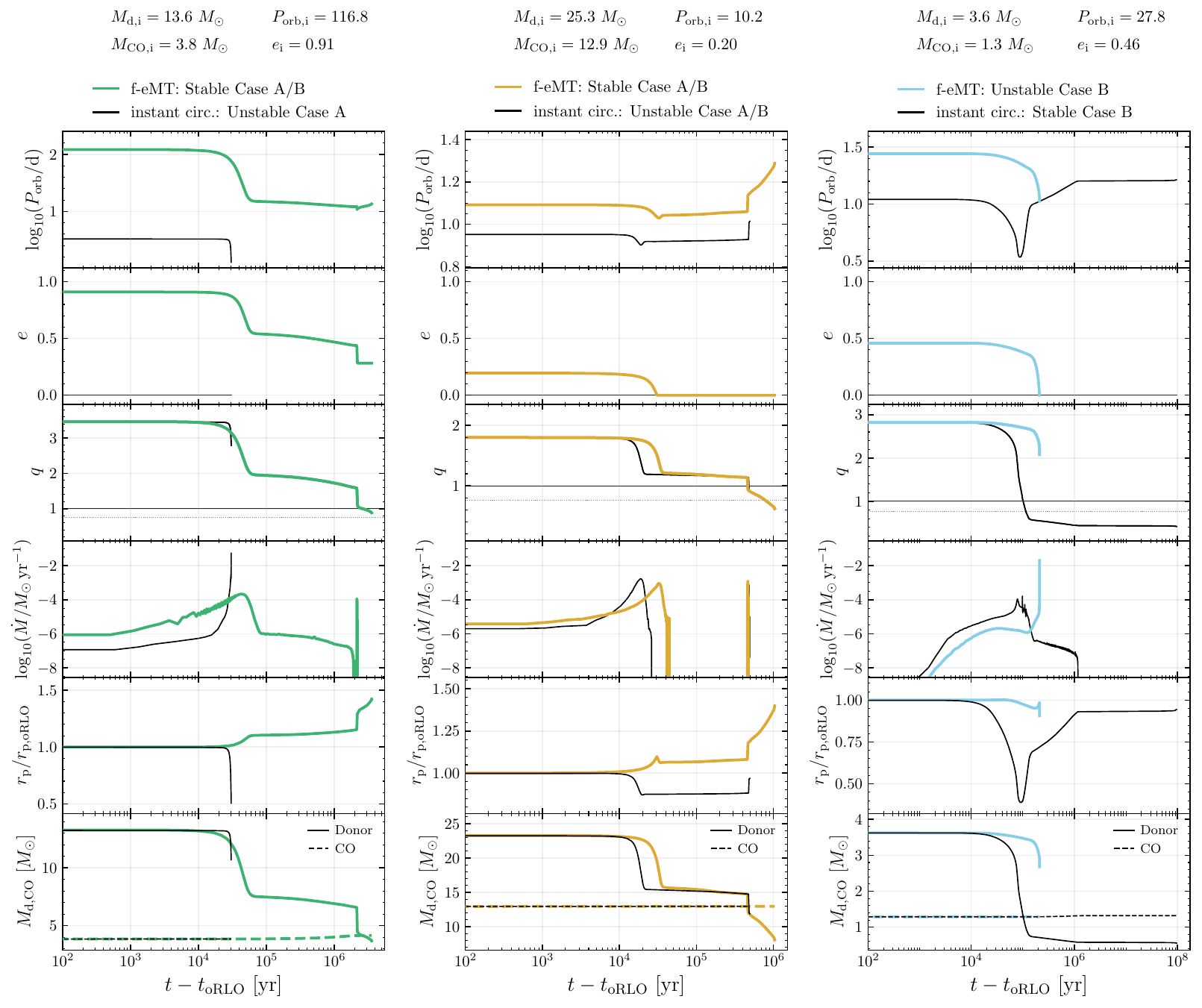}
    \caption{Time series evolution of three binaries which have divergent evolution between \ac{f-eMT} (colored lines) and their equivalent instantly circularized evolution (black) with initial conditions of the original eccentric orbit at the top of each column.
    Instantly circularized models are adjusted such that $a_\mathrm{c}=a(1-e)$.
    The majority of models which have divergent evolution are biased toward stable MT in the eccentric case ($72\%$), and unstable in the instantly circularized case ($21\%$). 
    In the first binary (left), the trend of increasing $r_\mathrm{p}$ allows the \ac{f-eMT} model to remain stable. 
    The second (center) binary naturally circularizes $e=0$ but the larger $P_\mathrm{orb,i}$ compared to the instant circularization allows the eccentric model to survive Case B MT.
    The last binary (right) becomes unstable once $e=0$ due to the lack of increasing $r_\mathrm{p}$ once this limit is hit, while the instantly circularized model transfers enough mass to avoid runaway MT.
    } 
    \label{fig:mismatch_timeseries}
\end{figure*}

\subsubsection{Divergent Evolution}\label{sec:results:divergent_evol}
In \autoref{fig:mismatch_timeseries}, we show three example binaries that exhibit completely divergent evolution (i.e. one model has stable MT while the other becomes dynamically unstable) when modeled with \ac{f-eMT} (colored lines) compared to the equivalent instantly circularized approximation (black). 
The \ac{f-eMT} simulations often exhibit lower peak $\dot{M}$ due to the increase in $r_\mathrm{p}$, while circular binaries always initially shrink when $q>1$.

The left column of \autoref{fig:mismatch_timeseries} shows a binary with a high mass donor ($M_\mathrm{d} \simeq 13\,M_\odot$) and a low-mass BH ($M_\mathrm{CO} \simeq 4\,M_\odot$) with high $e_\mathrm{i}\simeq 0.9$, undergoing stable Case A/B MT. 
In this case, the lower peak $\dot{M}$ in the \ac{f-eMT} model results in the BH accreting approximately $0.5\,M_\odot$. 
The corresponding instantly circularized binary evolves similarly until $\log_{10}(\dot{M}/M_\odot\,\mathrm{yr}^{-1}) \gtrsim -6$, at which point the shrinking separation drives a runaway increase in the MT rate.
The original binary with $P_\mathrm{orb,i} \simeq 10^2\,\mathrm{d}$ is significantly impacted by the instant circularization treatment, which arbitrarily reduces the initial orbit to $P_\mathrm{orb,circ} \simeq 3\,\mathrm{d}$.

The middle column of \autoref{fig:mismatch_timeseries} shows a binary hosting a massive donor $M_\mathrm{d}\simeq 25\,M_\odot$, but with a more massive \ac{BH} ($M_\mathrm{CO} \simeq 13\,M_\odot$) and lower $e_\mathrm{i} \simeq 0.2$. Shortly after Case A MT begins, the \ac{f-eMT} model naturally circularizes.
However, due to the larger orbital period of the \ac{f-eMT} model (since instant circularization reduces the semi-major axis, $a_\mathrm{c}=a(1-e)$), the post-MS Case B evolution (at $t-t_\mathrm{oRLO} \simeq 3 \times 10^5\,\mathrm{yr}$) is stable, whereas the instantly circularized model is deemed to enter dynamically unstable \ac{MT} due to $L_2$ overflow, which occurs when the donor star radius exceeds the second Lagrangian point.

The final column of \autoref{fig:mismatch_timeseries} shows a binary containing a low-mass donor $M_\mathrm{d}\simeq 3\,M_\odot$ and a $1.3\,M_\odot$ NS with moderate eccentricity ($e_\mathrm{i} \simeq 0.5$), where the \ac{f-eMT} simulation eventually becomes unstable during Case B MT, while the instantly circularized model remains stable.
The sharp increase in \ac{MT} near $\log_{10}(\dot{M}/M_\odot\,\mathrm{yr}^{-1}) \gtrsim -5$ occurs when the increase of $r_\mathrm{p}$ (see \autoref{sec:results:simple_case}) ceases abruptly at $e=0$, where further \ac{MT} rapidly shrinks the orbit in a runaway process.
Before this sharp spike in \ac{MT}, $r_\mathrm{p}$ decreases due to strong spin-orbit coupling, which spins up the donor.
The instantly circularized model artificially escapes a runaway \ac{MT} episode due to the higher average $\dot{M}$ throughout the \ac{MT} phase, reaching a low enough $q$ to switch from orbital shrinking to expansion.

\section{Discussion and Conclusions}
\label{sec:discussion_conclusions}
We have implemented the secular orbital evolution equations for non-conservative \ac{eMT} from \citealt{Sepinsky2009ApJ}, into self-consistent stellar evolution calculations for the first time using \mesa{}. Our focus is on binaries containing \acp{CO} and non-degenerate donors.
Our investigation is divided into two parts: (\textit{i}) a small parameter study with simplified physics where all orbital evolution is restricted to that induced by non-conservative \ac{eMT} (\autoref{sec:results:simple_case}); and (\textit{ii}) a full-physics grid with astrophysical initial conditions drawn from a \posydon{} BPS of \ac{CO}-hosting binaries that are found to initiate MT while eccentric (\autoref{sec:results:full_physics_models}). The key findings from this modeling are as follows:

\bigskip

1) \textbf{\ac{s-eMT} model:} In the \ac{s-eMT} simulations containing a $10~M_\odot$ BH and $20~M_\odot$ donor, we find novel behavior when including self-consistent modeling of the \ac{MT} rate from a \mesa{} stellar evolution model in combination with the orbital evolution due to \ac{eMT}.
For binaries with high $e_\mathrm{i}$, \ac{eMT} reduces $q$ before circularizing, allowing \adot{} and \edot{} to become positive. 
This causes the binaries to remain eccentric post-MT, with some systems experiencing eccentric pumping, where $e_\mathrm{f} > e_\mathrm{i}$.
Remarkably, while binaries are eccentric during \ac{MT}, the combined changes in \Porb{} and $e$ cause an effective \textit{increase} of $r_\mathrm{p}$. 
In contrast, the canonical circular case, $\dot{a}$ is negative for $q \gtrsim 1$.
This widening of the orbit tends to stabilize the binary to \ac{MT}, and in some cases leads to orders-of-magnitude lower \ac{MT} rates (compared to the circular case; up to a reduction of $\sim2.5$ dex in \autoref{fig:simple_models_time_series}), proceeding on the nuclear timescale for systems experiencing Case A \ac{MT}.
For the highly eccentric models, the lower \ac{MT} rate allows some \acp{BH} to accrete more conservatively, nearly doubling their mass before the donor's core-collapse episode.

\bigskip

2) \textbf{\ac{f-eMT} model:} The majority of evolutionary effects identified in our \ac{s-eMT} simulations also persist in our \ac{f-eMT} simulations.
The initial conditions of our \ac{f-eMT} simulations were drawn from an astrophysical BPS model, where $>70\%$ of all interacting CO-hosting binaries initiated \ac{eMT}, suggesting \ac{eMT} may strongly impact \ac{CO} binary populations.
Most binaries from our astrophysical population initiate post-MS \ac{MT} (Case B or Case C \ac{MT}), due to the wider orbital separations of binaries which experience \ac{eMT}.
We find that about one-third of binaries remain eccentric post-MT, while the remaining two-thirds naturally circularize during \ac{eMT}, with this division depending strongly on the assumed BPS physics.
While the eccentric post-MT population categorically diverges from the ad hoc treatment of instantaneous circularization (the ubiquitously adopted approximation), even models which naturally circularize through \ac{eMT} see notable differences in their final properties (with the $90\%$ distributions around $\delta M_\mathrm{d} \simeq -20\%$ and $\delta \Porb{} \simeq 100\%$).
A smaller fraction of binaries (about $5\%$ of the naturally circularized binaries) exhibit completely divergent evolution (compared to the instantly circularized approach), with \ac{eMT} generally predicting \ac{MT} stability over instability in $80\%$ of cases.

\bigskip

3) \textbf{A bifurcation in orbital evolution outcomes:} To determine which binaries are likely to remain eccentric post-MT or become naturally circularized, we find these outcomes are strikingly well separated in the $q_\mathrm{i}{-}e_\mathrm{i}$ plane. 
We have provided a fitting function in \autoref{eqn:q_e_polyfit} so that other studies (e.g. \ac{BPS} simulations) may estimate the impact of \ac{eMT} to first-order on different subpopulations such as gravitational wave sources and X-ray binary populations.
However, as we have shown, even models which are naturally circularized through \ac{eMT} can be significantly misrepresented by the ad hoc approach of instant circularization in the predicted donor masses and orbital configurations prior to SN or common-envelope.

\bigskip

4) \textbf{Behavior of systems containing a BH:} Approximately $60\%$ of all BH-hosting binaries in the \ac{f-eMT} models remain eccentric post-MT.
Due to their low $q_\mathrm{i}$, BH-hosting binaries are more likely to remain eccentric post-MT than NS-hosting binaries which predominantly circularize.
These BH-hosting binaries are also likely to undergo eccentric pumping and orbital expansion, with $\Porb{} \gtrsim 10^2$ days and $0.2 \lesssim e \lesssim 0.8$. 
This type of evolution may be relevant to some of the wide Gaia BH systems in recent discoveries \citep{El-Badry+2023MNRAS_BH1,El-Badry+2023MNRAS_BH2,GaiaCollab_Panuzzo+2024AA}, although our models would be relevant to BH companions that have been affected by \ac{MT}.
Future work incorporating \ac{eMT} into \mesa{} to simulate interactions between two non-degenerate stars will provide more direct comparisons to the wide Gaia binaries hosting \acp{CO}.

\bigskip

This investigation into the effects of \ac{eMT} on binary populations indicates that a wide range of \ac{CO} populations may be impacted by this new treatment.
For example, LMXB populations with wide circular orbits, or potential X-ray binary populations with eccentric and moderately wide orbits which may exhibit very low X-ray duty cycles (approximately $t_\mathrm{peri}/P_\mathrm{orb}$).
These binaries primarily initiate post-\ac{MS} \ac{MT}, leading to a short-lived X-ray transient stage followed by a detached quiescent phase with stripped helium donors at all eccentricities and wide periods ($10$ to $10^4$ days).
Given the selection effects that strongly bias against detecting wide binaries and helium stars \citep{Gotberg+2018AA,Drout+2023Sci}, it is unsurprising that no such binary has yet been discovered.
However, future studies with detailed models for the X-ray emission from binaries undergoing \ac{eMT} are necessary for drawing more robust conclusions.

In addition to XRBs, eccentric binaries containing BHs have been proposed to explain a special subclass of transient with late time non-thermal emission: fast luminous blue optical transients \citep[e.g.,][]{Lazzati+2024ApJ}.
Our results suggest an increased prevalence of these peculiar transients compared to standard binary stellar evolution models as we have shown that BH-binaries are more likely to remain eccentric post-MT.

One key assumption of our model for \ac{eMT} assumes all \ac{MT} occurs in a delta function at periapse.
Although hydrodynamical studies have shown the \ac{MT} rate in eccentric binaries resembles a Gaussian profile as a function of the true anomaly \citep[e.g. with FWHM $\sim0.12~\Porb{}$][]{Lajoie_Sills2011ApJ}, simulations often cover a single binary configuration and/or resolve too few orbits to determine secular orbital evolution \citep{Lajoie_Sills2011ApJ,Davis_Siess_Deschamps2013AA,Saladino_Pols2019AA}.
Analytic work extending beyond the assumption of delta function \ac{MT} are presently unable to account for non-conservative \ac{MT} \citep{Hamers_Dosopoulou2019ApJ}, which is important for binaries hosting \acp{CO} \citep[e.g.,]{Marino+2017AA,Ziolkowski_Zdziarski.2018MNRAS}.
Other secular effects such as those introduced by circumbinary disks from the aforementioned non-conservative accretion \citep[e.g.,][]{Valli+2024AA} or more detailed treatment of tides in eccentric orbits \citep[e.g.,][]{MacLeod2022ApJ,Sciarini2024AA} may lead to more complex behavior.

While including \ac{eMT} within the \posydon{} framework is our ultimate goal, simply including another dimension into our detailed binary grids is currently too computationally expensive.
Nevertheless, exploring more dimensions in binary stellar evolution such as eccentricity is important for our interpretation of modern astrophysical populations \citep[e.g.,][]{deSa+2024MNRAS.tmp.2312D}
Machine learning algorithms like active learning may be necessary to facilitate expansions to higher dimensions within the \posydon{} framework \citep[][]{Rocha2022ApJ}.

Combining all results from our \ac{f-eMT} simulations, we conclude \ac{eMT} is important to consider for the evolutionary history of \ac{CO}-hosting binaries (e.g. low-mass/high-mass X-ray binaries, binary black hole mergers, gamma-ray burst progenitors) which are thought to have at least one phase of \ac{RLO} \ac{MT}.
In future work we will implement the \ac{eMT} formalism into \mesa{} to simulate interactions between two non-degenerate stars.
In this regime, we plan to investigate the large population of wide post-interaction binaries observed to contain stripped donors and \acp{WD}.

\section*{Acknowledgments}
KAR, SG, and MS\ are supported by the Gordon and Betty Moore Foundation (PI Kalogera, grant awards GBMF8477 and GBMF12341) and CIERA's Riedel Family Fellowship in Data Sciece. 
DM and KAR\ thank the LSSTC Data Science Fellowship Program, which is funded by LSSTC, NSF Cybertraining Grant No.\ 1829740, the Brinson Foundation, and the Moore Foundation; their participation in the program has benefited this work.
KAR\ is also supported by the NASA grant awarded to the Illinois/NASA Space Grant Consortium, and any opinions, findings, conclusions, or recommendations expressed in this material are those of the author and do not necessarily reflect the views of NASA.
JJA acknowledges support for Program number (JWST-AR-04369.001-A) provided through a grant from the STScI under NASA contract NAS5-03127. 
VK was partially supported through the D.I.Linzer Distinguished University Professorship fund. 
S.B., M.B., T.F., M.K., and Z.X. acknowledge support by the Swiss National Science Foundation (PI Fragos, project number CRSII5$\_$213497).
KK and EZ were partially supported by the Federal Commission for Scholarships for Foreign Students for the Swiss Government Excellence Scholarship (ESKAS No.~2021.0277 and ESKAS No.~2019.0091, respectively).
KK is supported by a fellowship program at the Institute of Space Sciences (ICE-CSIC) funded by the program Unidad de Excelencia Mar\'ia de Maeztu CEX2020-001058-M.
Z.X. was supported by the Chinese Scholarship Council (CSC).
Support for M.Z. was provided by NASA through the NASA Hubble Fellowship grant HST-HF2-51474.001-A awarded by the Space Telescope Science Institute, which is operated by the Association of Universities for Research in Astronomy, Incorporated, under NASA contract NAS5-26555. 

The computations were performed at Northwestern University on the Trident computer cluster (funded by the GBMF8477 award). 
This research was supported in part through the computational resources and staff contributions provided for the Quest high performance computing facility at Northwestern University which is jointly supported by the Office of the Provost, the Office for Research, and Northwestern University Information Technology.
We thank Monica Gallegos-Garcia for assisting in the BPS setup for this work and Chase Kimball for helpful comments on the manuscript.

\software{\texttt{NumPy} \citep{numpy}, {\tt SciPy} \citep{scipy:2020NatMe..17..261V}, \texttt{Astropy} \citep{astropy:2013, astropy:2018, astropy:2022}, {\tt matplotlib} \citep{matplotlib}, {\tt pandas} \citep{reback2020pandas}, \posydon{} \citep{Fragos+2023ApJS,Andrews+2024arXiv}}, {\tt MESA} \citep{Paxton2011ApJS,Paxton2015ApJS,Paxton2013ApJS,Paxton2018ApJS,Paxton2019ApJS}

\bibliographystyle{aasjournal}
\bibliography{references.bib}

\appendix
\section{Supplementary Material}
\label{sec:appendix}
\begin{figure*}[ht]
    \centering
    \includegraphics[width=1\linewidth]{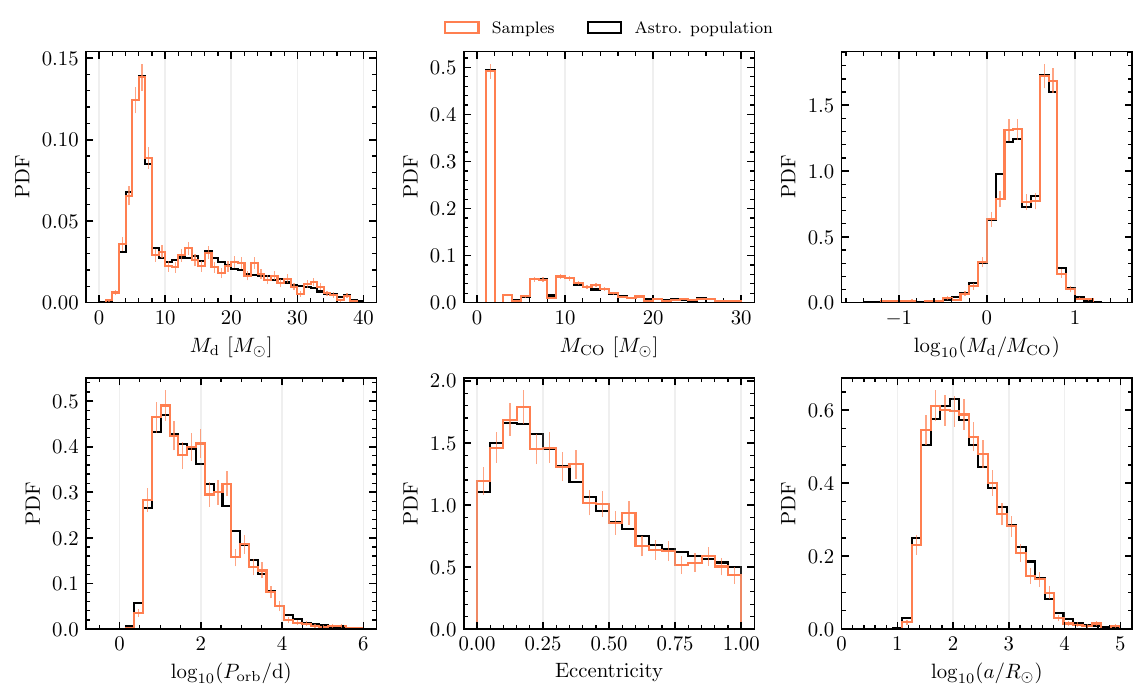}
    \caption{One-dimensional distributions of binary parameters from our \posydon{} binary population synthesis at the moment a non-degenerate donor star initiates \ac{RLO} with a BH or NS companion in an eccentric orbit ($e>0.05$; see Section \ref{sec:methods:popsyn_initial_conditions}). 
    The astrophysical binary population (black) is well reproduced by the subset of $2000$ samples we simulate with \ac{f-eMT} in \mesa{} (orange) where vertical lines indicate Poisson errors.} 
    \label{fig:astro_pop_to_MESA}
\end{figure*}

\restartappendixnumbering
To obtain realistic initial conditions for our \ac{eMT} simulations with binaries hosting a BH or NS, we take properties from a \posydon{} \ac{BPS} of systems which initiate \ac{MT} in a nominally eccentric orbit ($e>0.05$).
In \autoref{fig:astro_pop_to_MESA}, we show this binary population in black, while the samples used for running our \ac{f-eMT} \mesa{} simulations are shown in orange with Poisson error bars.
Error bars on the astrophysical population would be too smaller than the line width.
The binary evolution physics prior to oRLO (MT, SN, common envelope evolution, tides) will impact the inferred distribution of binaries which initiate \ac{eMT}.
Other uncertainties in binary evolution, such as the birth eccentricity distribution of MS binaries and the initial mass function will also impact this population.

In our \ac{f-eMT} simulations, we find a population of binaries which naturally circularize during \ac{eMT}. 
We then compare these systems to \mesa{} simulations run with the instantly circularized configuration \autoref{sec:methods:inst_circ}.
The data in \autoref{table:MT_cases_circularized} summarize the MT outcomes of our simulations, comparing the \ac{f-eMT} models with the instantly circularized ones.

\begin{table}[]
    \caption{Mass transfer outcomes for our \ac{f-eMT} \mesa{} simulations run with the eccentric MT formalism ($e_\mathrm{i} \neq 0$), and their associated instantly circularized simulations. 
    This table only shows models which circularize naturally through \ac{RLO} \ac{MT} ($e_\mathrm{f} < 0.05$), which constitute about $2/3$ of all our astrophysically sampled models (see \S\ref{sec:results:natty}. 
    For each simulation pair, we show the top three most numerous joint MT cases (reported fractions may not sum to unity).
    A large fraction ($96\%$) of all models agree in their MT histories, with $4\%$ of models exhibiting completely divergent evolution (i.e. stable vs. unstable MT). 
    For models which agree on the qualitative outcome, they also see good agreement between their detailed MT cases (e.g. both Case B).}\label{table:MT_cases_circularized}
    \begin{tabular}{c|c|c|c|c}
    Outcome                    & N models                  & Eccentric MT  ($e_\mathrm{f}<0.05$) & Instantaneous Circ.  ($e_\mathrm{i}=0$) & Fraction \\ \hline \hline
    \multirow{3}{*}{Stable MT} & \multirow{3}{*}{328 (28\%)} & Case B          & Case B               & 52.8\% \\
                               &                             & Case A/B         & Case A/B               & 45.6\% \\
                               &                             & Case B          & Case A/B          & 0.6\%  \\ \hline
    \multirow{3}{*}{Unstable}  & \multirow{3}{*}{790 (68\%)} & Case B          & Case B               & 91.9\% \\
                               &                             & Case A          & Case A               & 7.4\%  \\
                               &                             & Case B          & Case A           & 0.4\%  \\ \hline
    \multirow{3}{*}{Mixed TFs} & \multirow{3}{*}{47 (4\%)} & Stable Case B \& Case B/BB        & Unstable Case B                         & 64.1\%   \\
                               &                             & Stable Case A/B  & Unstable Case A \& Case A/B & 14.1\% \\
                               &                             & Unstable Case B & Stable Case B    & 12.5\%

    \end{tabular}
\end{table}

\end{document}